\documentclass{jfm}
\pdfoutput=1 
\usepackage{graphicx}  
\usepackage{epstopdf,epsfig}
\usepackage{textgreek}
\usepackage{amsmath}
\usepackage{subfigure}\usepackage{float}\usepackage{epsfig}
\usepackage{amssymb}\usepackage{psfrag}
\usepackage{stmaryrd}
\usepackage{picture}
\usepackage{float}
\usepackage{natbib}
\usepackage{hyperref}
\hypersetup{colorlinks=true,breaklinks=true,linkcolor=blue}
\usepackage{multirow}
\newcommand{\be}{\begin{equation}}
\newcommand{\ee}{\end{equation}}
\newcommand{\bef}{\begin{figure}}
\newcommand{\ef}{\end{figure}}
\newcommand{\bear}{\begin{eqnarray}}
\newcommand{\ear}{\end{eqnarray}}
\newcommand{\barr}{\begin{array}}
\newcommand{\earr}{\end{array}}

\newcommand{\nn}{\nonumber}
\newcommand{\td}[1]{\tilde{#1}}
\newcommand{\f}{\frac}

\newcommand{\del}{\partial}
\newcommand{\bs}{\boldsymbol}

\newcommand{\msb}{\mathbf}

\shortauthor{Nehal Dash and Ganesh Tamadapu}

\title{Radial dynamics of an encapsulated microbubble with interface energy}

\author{N. Dash\aff{} \and G. Tamadapu\aff{}
  \corresp{\email{gt@iitm.ac.in}}}

\affiliation{
  Department of Applied Mechanics, Indian Institute of Technology Madras,
Chennai, 600036, India.}
\begin{document}
\maketitle

\begin{abstract}
In this work, a mathematical model based on interface energy is proposed within the framework of surface continuum mechanics to study the dynamics of encapsulated bubbles. The interface model naturally induces a residual stress field into the bulk of the bubble, with possible expansion/shrinkage from a stress-free configuration to a natural equilibrium configuration. The significant influence of interface area strain and the coupled effect of stretch and curvature is observed in the numerical simulations based on constrained optimization. Due to the bending rigidity related to additional terms, the dynamic interface tension can become negative, but not due to the interface area strain. The coupled effect of interface strain and curvature term observed is new and plays a dominant role in the dominant compression behavior of encapsulated bubbles observed in the experiments. We validate our model by fitting experimental data of $1.7\,\mu$m, $1.4\,\mu$m, and $1\,\mu$m radii bubbles by calculating the optimized parameters. Our work also highlights the role of interface parameters and natural configuration gas pressure in estimating the size-independent viscoelastic material properties of encapsulated bubbles with interesting future developments.
\end{abstract}

\maketitle
\section{Introduction}
Coated or encapsulated bubbles (EBs) have a large number of applications in the field of medical science, for targeted drug delivery \citep{1347-4065-38-5S-3014,UNGER200145,Tsutsui2004,UNGER20041291,LIU200689,HERNOT20081153,KOOIMAN201428}, as contrast agents for ultrasound medical imaging \citep{hoff2001acoustic,Stride2003,Lindner2004,postema2004ultrasound,Klibanov2006} and so on. Though EBs are widely used for ultrasound medical imaging, in particular, a recent study \citet{errico2015ultrafast} have shown that these bubbles can not only be used as rough contrast agents, rather can also be used as traces to extract vascular geometry (microvascular imaging) and blood flow velocity over a large dynamic range in microns length scales. The encapsulation of such bubbles can be made up of polymers \citep{Liu2014,Song2018}, proteins \citep{Wang2020} or phospholipids \citep{BuchnerSantos2012,Li2013,AbouSaleh2013,Gong2014,Parrales2014,vanRooij2015,Segers2016,Lum2016,Helfield2019,Shafi2019,Segers2020}. Thus experimental and theoretical understanding of the response of EBs to the external acoustic field is of great importance for the future developments of these promising applications/techniques.

Many experimental studies had been carried out to understand the behavior of the EBs where different techniques have been used for oscillating the EBs. \citet{DEJONG199295} and \citet{DEJONG1993175} used bubble scattering phenomenon and mentioned that the scattering, absorption, and extinction cross-section are vital acoustic properties of EBs, especially in echocardiographic applications. Many models were developed for describing the acoustic scatter and attenuation from suspensions of the EBs \citep{FRINKING1998523,584337}. Taking a step further, \citet{Morgan2000} considered the echoes from wideband insonation, the effect of transmitted phase, and further compared with theoretical predictions by experimentally measuring the scattered pressure from a single EB. \cite{GORCE2000} performed a study on SonoVue\textsuperscript{\textregistered} and determined the shell parameters from a comparison study of the model with that of experimental measurements of the attenuation coefficients/expressions. Experiments have also been carried out following the acoustic spectroscopy approach by transmitting a sequence of tone bursts within a range of frequencies \citep{vanderMeer2007,Helfield2013}. \citet{Tu2009} performed experiments using light scattering methods to study the dynamical response of the EBs when subjected to an ultrasound field. Furthermore, they considered three different and popular shelled bubble dynamics models \citep{DEJONG1993175,sarkar2005,marmottant2005} and tried to comment on the best bubble shelled model by fitting the experimental data. They estimated the shell parameters by fitting the experimental light scattering data with the linearized version of the Marmottant model and discussed the variation of shell elasticity modulus $(\chi)$ and shell dilatational viscosity $(k^{\rm S})$ with the initial radii of the bubble. In a detailed review, \citet{VERSLUIS20202117} have highlighted the development of the EBs along with various experimental measuring methods.

For the theoretically developed EB models, one of the critical aspects is to inspect if the theoretical model is in good agreement with the experimental observations/fittings. It could be fitting the experimental time series response or estimating the shell properties such as shell elasticity modulus $(\chi)$, shell dilatational viscosity $(k^{\rm S})$ and so on, though with some additional assumptions. Assuming the encapsulation as a thin shell, various researchers have developed mathematical models for contrast agent microbubbles. Complementing such shell models are the studies of \cite{DEJONG199295,DEJONG1993175,DEJONG1994447,Church1995}, who came up with a model by adding elastic and viscous terms of the shell into the Rayleigh-Plesset (RP) equation. Shell model was also developed considering zero thickness interface bubble, assuming that the encapsulation behaves similarly to that of a Newtonian fluid \citep{sarkar2005}. Dynamic simulation of lipid shelled microbubbles was carried out by \citet{marmottant2005}, referred to as the Marmottant model. These authors have studied the interface tension in an ad hoc manner, which is empirical but does not provide any relevant physical phenomenon. To resolve this, \citet{Paul2010} considered interface strain-dependent surface tension with an interface elasticity constant. In the process of fitting the experimental data, these models \citep{marmottant2005, Paul2010} assumed that some material properties of the encapsulation depend on initial bubble size. 
Models were also developed with the nonlinear theory for shell viscosity \citep{DOINIKOV2009263}, which depicted the dominant compression behavior in encapsulated microbubbles. Some models \citep{Qin2010} considered the shell elastic term valid for finite deformations, unlike Church model \citep{Church1995}, which is valid only for small deformations. \cite{tsiglifis2008nonlinear} developed an EB model, where the encapsulating shell was modeled as a thin membrane following hyperelastic constitutive laws. This model emphasized the flow structure aspect of radial bubble dynamics. Recently, \citet{doi:10.1121/10.0003500} developed an EB model where the bubble shell is considered to be a compressible viscoelastic isotropic material and was later generalized to an anisotropic material.

Though many EB models are limited to the study of radial oscillations, for an EB in an ultrasound field, the shape instabilities are equally important. Using optical imaging, such instabilities were even observed in experiments demonstrating the destruction of the bubble during compression \citep{chomas2000optical}. Motivated by these observations, many EB models were proposed to describe the non-spherical oscillations of EB dynamics \citep{1417703,4151897,dollet2008nonspherical,PhysRevE.82.026321,vos2011nonspherical}.

\citet{steigmann} developed a theory for coupled three dimensional deformations of elastic solids with elastic films at their bounding surfaces. They proposed a simple theoretical derivation validating finite deformations and compatible with general mechanical principles. They assumed the film to be an elastic surface that resists the variations in its metric and curvature. This approach further opened ways to solve various complex technical problems, which included any isotropic substrates with hemitropic or isotropic films attached to the surfaces using the generalized kinematics of surfaces. Following the work of \citet{steigmann} and based on the results from other theoretical, experimental and computational studies explaining/supporting the size-dependence of interface energy and interface stresses \citep{tolman1949effect,jiang2008size,medasani2009computational}, \citet{GAO201459} proposed an interface theory which broadly explained the curvature dependence of interface energy and the effect of the residual elastic field on the bulk due to the interface energy. They described the deformation due to interface energy by hypothetically splitting the solid into homogeneous pieces/splits along the interface. They also emphasized the fact that this splitting is purely an imaginary action and hence named it as ``fictitious stress-free configuration", but it provided a strong/valid explanation for the calculation of residual elastic field due to the interface energy.

It is understandable that when an EB is suspended in a fluid, the interfaces between the gas-encapsulation and encapsulation-liquid carry sufficient surface/interface energy that significantly affects the mechanics. The influence of this interface energy on EB dynamics is not well studied in the existing models. Although models with interface strain are considered \citep{marmottant2005,Paul2010} by directly adding it to the interface tension, these models do not capture the essential interface mechanics due to the lack of a proper continuum framework. However, recent literature shows the importance of deformation dependent surface/interface energy of polymeric glass, thin films, and networks \citep{ACSletter,PRL1,Nat1}, and the importance of interface stretch ratio and curvature effects in the liquid-vapor interface and growing droplet \citep{PRL2003,PRL2016}. It was reported that when the polymer network attains maximum stretch, the strain dependent surface energy shows its significance in terms of surface stresses \citep{ACSletter}. Despite various shell models developed in the literature, a better understanding of EB dynamics is lacking, and a thermodynamically consistent model including interface effects is required to explain the physical origins of EB characteristic behavior observed in the experiments (using high-speed imaging techniques). One such important characteristic is the ``compression-only" behavior where the EB compresses significantly compared to its expansion.

In this paper, we try to address these questions by proposing an interface energy model for the dynamics of gas filled EB suspended in a fluid. As we are keener on studying the behavior of EBs with a few microns radius, the effect/influence/significance of interface energy becomes more dominant. Notably, the present model considers these interface effects and makes this a robust model over the existing literature. Motivated by the general framework of \citet{steigmann}, we assume that the interfaces are of {\it Steigmann-Ogden} type and calculate the additional terms from the modified Young-Laplace equation for the radial dynamics of EB. For a bubble of $2\,\mu$m inner radius with $20\,$nm thickness, we estimate the interface and usual material parameters at a particular excitation pressure and frequency by constructing a nonlinear constrained minimization problem for dominant compression behavior. We further analyze various bubble configurations and study the contributions of the parameters introduced in the present model. The results show that interface stretch and bending rigidity induced bulk residual stress plays a dominant role in the mechanics of EBs with possible negative dynamic inner interface tension. Later, we attempt to compare the present model for $1.7\,\mu$m, $1.4\,\mu$m, and  $1\,\mu$m radii bubbles with existing experimental data and identify the importance of interface parameters introduced in this work. With the proposed model, we show that size-independent shell viscoelastic properties can be estimated for \cite{Tu2009} experimental data.

This paper is organized as follows. In Section~\ref{sec:Mathmodel}, a detailed mathematical model is presented for the radial dynamics of EB. In Section~\ref{sec:IEM}, we introduce the concept of fictitious configuration through the interface energy model and obtain the additional terms in the interface tension due to the modified Young-Laplace equation. In Section~\ref{sec:cr}, we discuss the constitutive relations of the bulk and the fluid. In Section~\ref{sec:ge}, we obtain the governing equation for the radial dynamics of EB as a modified Rayleigh-Plesset equation with additional terms due to the interface energy. In Section~\ref{sec:co}, we introduce a constrained optimization procedure to estimate the interface and material parameters and present the numerical results and validation with experiments in Section~\ref{sec:results}. The conclusions are given in Section~\ref{sec:conclusion}.

\section{Mathematical Model}
\label{sec:Mathmodel}

An important step in modeling multiphase interfaces is to identify the surface/interface excess thermodynamic variables along with that of the bulk phase variables \citep{Sagis2014}. One such explicit important variable is the interface curvature. In fact, interface curvature can have a significant implicit influence on other thermodynamic variables such as interface energy, chemical potential, and surface charge density. We first present an interface energy model based on the interface strain and curvature in the following sections. Later, we obtain the governing differential equation for the radial dynamics of EB with additional terms due to the interface energy.

\subsection{Interface energy model}
\label{sec:IEM}
\textcolor{black}{In the existing literature \citep{helfrich1973elastic,Rangamani2013,ARGUDO20161619,bian2020bending}, modeling of cell membranes and lipid bi-layers is performed using Helfrich curvature energy which takes care of the bending energy through Gaussian curvature and spontaneous curvature. However, one cannot directly use Helfrich energy to model interfaces, which requires a generic framework to identify the structure of interface energy in terms of material symmetries. Also, it is important to note that these models are used to study the initial natural states of micro/nanoscale structures. On the other hand, we are interested in modeling encapsulated microbubbles with the effect of residual fields developed due to the interface energy. Therefore, we follow the generic continuum framework developed in the previous work \citep{GAO201459,steigmann,yuan-chengfung1977} to incorporate the interface energy effects.} 

\begin{figure}
\centering
\includegraphics[width=0.9\linewidth]{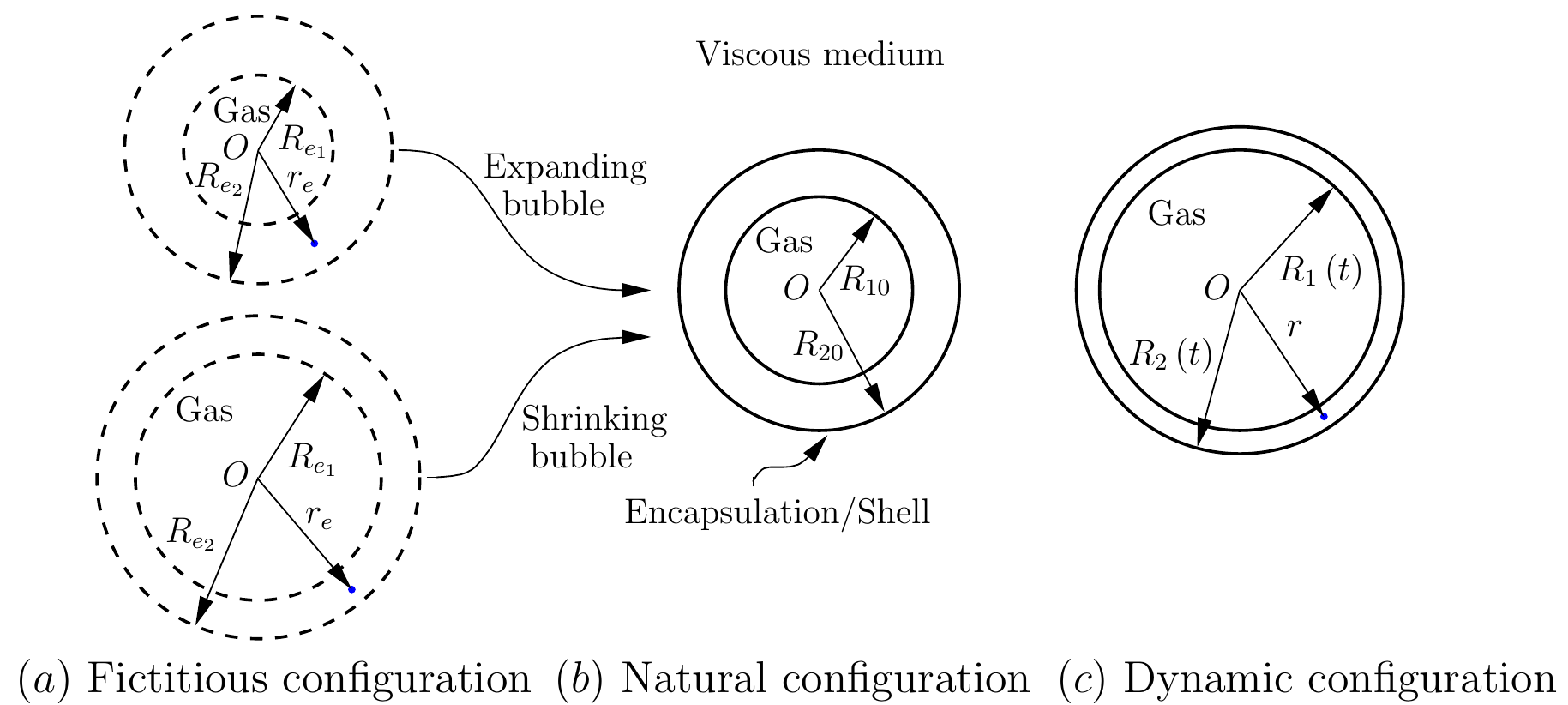}
\caption{Schematic representation of a gas filled encapsulated microbubble suspended in a viscous medium at three different configurations. (\textit{a}) Fictitious (bulk stress free)  configuration (FC) of the bubble with possible expansion/shrinkage to reach the natural (static equilibrium) configuration (NC) in (\textit{b}) and (\textit{c}) Dynamic configuration (DC) of the bubble at an instant of time $t$.}
\label{f:bubble}
\end{figure}

We assume that the bubble interface is a deformable mathematical surface compatible with the deformation of the bulk. We assume that there exist a bulk stress-free fictitious configuration (FC) of the bubble as shown in figure~\ref{f:bubble}(\textit{a}) with inner and outer radii, $R_{e_1}$ and $R_{e_2}$, respectively. Even in the absence of external load this FC is unstable due to the presence of excess interface energy and the corresponding non-zero interface stress, which can force the bubble to attain a stable self-equilibrium state through expansion/shrinkage. This stable self-equilibrium state with bulk residual stress developed due to heterogeneous interface energy is shown in figure~\ref{f:bubble}(\textit{b}) with inner and outer radii $R_{10}$ and $R_{20}$, respectively. We also call this configuration as natural configuration (NC). Therefore, the bulk residual stress field in the bubble can be described as the state of self-equilibrium stress field due to the residual interface stresses. When excited by an external ultrasound pressure field, figure~\ref{f:bubble}(\textit{c}) represents the radial dynamic configuration (DC) of the same bubble with time dependent inner and outer radii, $R_1(t)$ and $R_2(t)$, respectively.

The approach presented above allows accounting for the influence of excess interfacial free energy and the corresponding non-zero residual interface stress on the mechanical behavior of EBs. This FC is in agreement with the imaginary ``fictitious stress-free configuration" suggested/reported by \citet{GAO201459}. It is important to note that this FC may not exist physically. However, this assumption will help in calculating the bulk residual stress field induced by interfacial energy. Though many researchers have ignored the role of interfacial energy induced bulk residual stress, it should be noted that when we consider the effect of interface energy, the bulk residual stress field induced by the interfacial energy always becomes an important characteristic to account for and hence cannot be ignored. In very recent work, \citet{doi:10.1121/10.0003500} have also highlighted the importance of this stress-free configuration for developing dynamic models for EBs.

Let $(x^{\rm1},x^{\rm2},x{\rm^3})$ be the three dimensional Cartesian coordinate system with unit basis vector triad as  $(\bs{e}_1,\bs{e}_2,\bs{e}_3)$. Let $(\rho,\phi,\theta)$ be the  principal spherical polar coordinates with unit basis vector triad $(\bs{e}_\rho,\bs{e}_\phi,\bs{e}_\theta)$. Here, $\rho$ is the radial coordinate from the centre of the bubble, $\phi$ is the polar angle formed with $x^{3}-$axis and $\theta$ is the azimuthal angle measured about $x^3-$axis in the counter clockwise direction from $x^1-$axis. The relation between the Cartesian and the spherical polar coordinates is given by
\begin{align}
\begin{split}
    x^{1}&=\rho\sin\phi\cos\theta,\\
    x^{2}&=\rho\sin\phi\sin\theta,\\
    x^{3}&=\rho\cos\phi.
\end{split}
\end{align}
The unit vectors along the three spherical coordinate directions are given by
\begin{align}
\begin{split}
    \bs{e}_\rho&=\sin\phi\cos\theta\,\bs{e}_1+\sin\phi\sin\theta\,\bs{e}_2+\cos\phi\,\bs{e}_3,\\
    \bs{e}_\phi&=\cos\phi\cos\theta\,\bs{e}_1+\cos\phi\sin\theta\,\bs{e}_2-\sin\phi\,\bs{e}_3,\\
    \bs{e}_\theta&=-\sin\theta\,\bs{e}_1+\cos\theta\,\bs{e}_2.
\end{split}
\end{align}
For describing bulk radial motion of the bubble, we assume $\bs{x}_e=r_e \bs{e}_\rho$ as the position vector of a material point in FC with radius $r_e$. Similarly, let $\bs{x}=r(r_e,t)\bs{e}_\rho$ be the position vector of a material point in DC. Then the bulk deformation gradient tensor, ${\mathbf A}$, and the corresponding left Cauchy-Green deformation tensor, ${\mathbf B}$, can be expressed as \citep{itskov2007tensor}
\begin{align}
\begin{split}
{\mathbf A}&={\partial\bs{x}}/{\partial \bs{x}_e}=\frac{\partial{r}}{\partial r_e}\bs{e}_\rho\otimes\bs{e}_\rho+\frac{r}{r_e}\bs{e}_\phi\otimes\bs{e}_\phi+\frac{r}{r_e}\bs{e}_\theta\otimes\bs{e}_\theta,\\
{\mathbf B}&=\msb{A}\msb{A}^{\rm T}=\left(\frac{\partial{r}}{\partial r_e}\right)^2\bs{e}_\rho\otimes\bs{e}_\rho+\left(\frac{r}{r_e}\right)^2\bs{e}_\phi\otimes\bs{e}_\phi+\left(\frac{r}{r_e}\right)^2\bs{e}_\theta\otimes\bs{e}_\theta,
\end{split}
\end{align}
where $\bs{c}\otimes\bs{d}$ represents the tensor product of two vectors $\bs{c}$ and $\bs{d}$. 

Following the framework of \cite{steigmann}, we idealize the gas-encapsulation and encapsulation-liquid interfaces as mathematical interfaces of zero thickness and call it Steigmann-Ogden interfaces (SOIs). Let $\theta^1=\phi$ and $\theta^2=\theta$ be the surface coordinates of the spherical interface of radius say $r_0$. In the convected coordinate system, let $r(r_0,t)$ be the deformed radius of the interface. For the bubble undergoing spherical deformation, the principal stretch ratio $\lambda=r(r_0,t)/r_0$.
Let $\bs Z(\theta^1,\theta^2)$ and $\bs z(\theta^1,\theta^2)$ be the position vectors of the same point on the undeformed and deformed interfaces, respectively, given by 
\begin{align}
\begin{split}
    \bs Z(\theta^1,\theta^2)&=r_0(\sin\theta^1\cos\theta^2\,\bs{e}_1+\sin\theta^1\sin\theta^2\,\bs{e}_2+\cos\theta^1\,\bs{e}_3),\\
    \bs z(\theta^1,\theta^2)&=r(r_0,t)(\sin\theta^1\cos\theta^2\,\bs{e}_1+\sin\theta^1\sin\theta^2\,\bs{e}_2+\cos\theta^1\,\bs{e}_3).
\end{split}
\end{align}
Due to the deformation of the bulk of the bubble, the interface can be assumed to be convected and the deformation mapping $\bs \chi$, relates the same point before and after deformation as follows 
\begin{equation}
\bs z(\theta^1,\theta^2)=\bs \chi(\bs Z(\theta^1,\theta^2)).
\end{equation}
For the given position vectors of the interface, the tangent vectors $\bs G_\alpha$ and $\bs g_\alpha$ on the undeformed and deformed interfaces, respectively, are calculated using 
\begin{align}
\begin{split}
    \bs{G}_\alpha=\bs{Z}_{,\alpha},\quad
    \bs{g}_\alpha=\bs{z}_{,\alpha},\quad
    \alpha\in\{1,2\},\quad
    (\cdot)_{,\alpha}=\f{\del(\cdot)}{\del\theta^\alpha},
\label{e:tgtvec}
\end{split}
\end{align}
and the expressions for these vectors are given by
\begin{alignat}{4}
    \bs{G}_1&=r_0\,\bs{e}_\phi,\quad
    \bs{G}_2=r_0\sin\phi\,\bs{e}_\theta, \\
    \bs{g}_1&=r(r_0,t)\,\bs{e}_\phi,\quad
    \bs{g}_2=r(r_0,t)\sin\phi\,\bs{e}_\theta.
\end{alignat}
The components of the interface metric tensor for the undeformed and deformed interfaces, respectively, are given by
\begin{alignat}{2}
    G_{\alpha\beta}&=\begin{pmatrix}
                    \bs{G}_1\cdot\bs{G}_1 & \bs{G}_1\cdot\bs{G}_2\\
                    \bs{G}_2\cdot\bs{G}_1 & \bs{G}_2\cdot\bs{G}_2
                    \end{pmatrix}=r_0^2\;\rm{diag}(1,\sin^2\phi),
                    \label{e:Gdown}\\
    g_{\alpha\beta}&=\begin{pmatrix}
                    \bs{g}_1\cdot\bs{g}_1 & \bs{g}_1\cdot\bs{g}_2\\
                    \bs{g}_2\cdot\bs{g}_1 & \bs{g}_2\cdot\bs{g}_2
                    \end{pmatrix}=r(r_0,t)^2\;\rm{diag}(1,\sin^2\phi) =\lambda^2 \it{G}_{\alpha\beta}.
                    \label{e:gdown}
\end{alignat}
For the given undeformed metric tensor components $G_{\alpha\beta}$, the Christoffel symbols are defined as
\begin{align}
    \td{\Gamma}_{\;\,\alpha\beta}^\gamma=\f{1}{2}G^{\gamma\delta}(G_{\alpha\delta,\beta}+G_{\delta\beta,\alpha}-G_{\beta\alpha,\delta}).
\end{align}
Similarly, $\Gamma_{\;\,\alpha\beta}^\gamma$ are the Christoffel symbols constructed using the deformed interface metric tensor components $g_{\alpha\beta}$. The covariant derivative of a tensor generalizes the concept of partial derivative onto the curved surface. The following relation obtains the calculation of the components of covariant derivative $(:)$ of a vector $(v^\alpha)$
\begin{align}
    {v^\alpha}_{:\beta}={v^\alpha}_{,\beta}+\Gamma^\alpha_{\;\,\gamma\beta}\,v^\gamma.
\end{align}
The dual tangents vectors represented by $\bs{G}^\alpha$ and $\bs{g}^\alpha$ associated with the coordinates $\theta^\alpha$ are given by
\begin{align}
\begin{split}
        \bs{G}^\alpha=G^{\alpha\beta}\bs{G}_\beta,\quad
        \bs{g}^\alpha=g^{\alpha\beta}\bs{g}_\beta.
\label{e:dtgtvec}
\end{split}
\end{align}
The calculated values of the components of the dual metric tensors $G^{\alpha\beta}$ and $g^{\alpha\beta}$, respectively, are obtained as
\begin{align}
    G^{\alpha\beta}&=r_0^2\;\rm{diag}(1,\csc^2\phi),\label{e:Gup}\\
    g^{\alpha\beta}&=r(r_0,t)^2\;\rm\rm{diag}(1,\csc^2\phi)=\lambda^{-2}\,\it{G}^{\alpha\beta}.\label{e:gup}
\end{align}
The second fundamental forms $Q_{\alpha\beta}$ and $q_{\alpha\beta}$ representing the normal curvatures associated with the deformed and the undeformed surfaces, respectively, are given by
\begin{align}
\begin{split}
    Q_{\alpha\beta}=\bs{N}\cdot\bs{G}_{\alpha:\beta},\quad q_{\alpha\beta}=\bs{n}\cdot\bs{g}_{\alpha:\beta}.
\label{e:qtensor}
\end{split} 
\end{align}
Here, $\bs{N}$ and $\bs{n}$ are the outward normal vectors to the undeformed and deformed interfaces, respectively, given by
\begin{align}
\begin{split}
    \bs{N}&=\frac{1}{2}\mu^{\alpha\beta}\bs{G}_\alpha\times\bs{G}_\beta,\\
    \bs{n}&=\frac{1}{2}\varepsilon^{\alpha\beta}\bs{g}_\alpha\times\bs{g}_\beta,\quad\text{(summation convention)}
\label{e:normal}
\end{split}
\end{align}
where $\varepsilon^{\alpha\beta}=e^{\alpha\beta}/\sqrt{g}, \mu^{\alpha\beta}=e^{\alpha\beta}/\sqrt{G}$ with $G={\rm det}(G_{\alpha\beta})=r_0^4\sin^2\phi, g={\rm det}(g_{\alpha\beta})=r(r_0,t)^4\sin^2\phi$ and $e^{\alpha\beta}=e_{\alpha\beta}$ is the alternator symbol ($e^{12}=-e^{21}=1, e^{11}=e^{22}=0$). Further we calculate the following relations
\begin{align}
    \bs{G}_{1,1}=-r_0\,\bs{e}_r,\quad \bs{G}_{1,2}=\bs{G}_{2,1}=r_0\cos\phi\,\bs{e}_\theta, \quad
    \bs{G}_{2,2}=-r_0\sin\phi\,(\cos\theta\,\bs{e}_1+\sin\theta\,\bs{e}_2),
\label{e:Gg}
\end{align}
and obtain the expression for $Q_{\alpha\beta}$ using the relations in \eqref{e:qtensor} and \eqref{e:normal} where $\bs{N}=\bs{e}_\phi \times \bs{e}_\theta=\bs{e}_\rho$ as follows
\begin{align}
    Q_{\alpha\beta}=-r_0\;\rm{diag}(1,\sin^2\phi)
    =-{\it{r}}_0^{-1}\it{G}_{\alpha\beta}.
\label{e:Qab}
\end{align}
From the kinematics of the interface deformation, the deformation gradient tensor $\bf{a}$ can be written as \citep{itskov2007tensor}
\begin{align}
    \bf{a}=\bs{g}_\beta\otimes\bs{G}^\beta.
\end{align}
The right Cauchy green deformation tensor ${\msb C}$ is isotropic and can be expressed as
\begin{align}
    \msb{C}&=\msb{a}^{\rm T}\msb{a}=g_{\alpha\beta}\bs{G}^\alpha\otimes\bs{G}^\beta.
\end{align}
The curvature tensor $(\bf{q})$ and the relative curvature tensor $(\bs{\kappa})$, related to the deformed and the undeformed configurations, respectively, are defined as 
\begin{align}
\begin{split}
    \msb{q}&=q_{\alpha\beta}\,\bs{g}^\alpha\otimes\bs{g}^\beta,\\
    \bs\kappa&=\kappa_{\alpha\beta}\,\bs{G}^\alpha\otimes\bs{G}^\beta,\\
\end{split}
\end{align}
with the relation $\bs\kappa=-\msb{a}^{\rm T}\msb{q}\,\msb{a}$. Then the relation between the components of the above two curvature tensors is $\kappa_{\alpha\beta}=-q_{\alpha\beta}$,
where $q_{\alpha\beta}$ is calculated using equation \eqref{e:qtensor}.

In a detailed study accounting for the material symmetry of the interface, \citet{steigmann_2001} concluded that the interfacial energy is not an isotropic scalar-valued tensor function corresponding to the reference configuration, rather it is a function of right Cauchy-Green interface deformation tensor $\msb{C}$ and the relative curvature tensor $\bs{\kappa}$. Here, the interface is assumed to be a micropolar,  which is not isotropic with respect to inversion, also known as hemitropic. For such hemitropic interface, its energy density is expressed as \citep{steigmann,yuan-chengfung1977,GAO201459} 
$$\gamma=\gamma(\msb{C},{\bs \kappa}).$$
Hence, the interface energy density can be expressed in terms of six basis invariants of the right Cauchy-Green interface deformation tensor $\msb{C}$ and the relative curvature tensor $\bs{\kappa}$ as 
$$\gamma=\gamma(I_1, I_2, I_3, I_4, I_5, I_6),$$ 
and the invariants are given by
\begin{align}
\begin{split}
    I_1&={\rm tr}\,\msb{C}=G^{\alpha\beta}C_{\alpha\beta},\\
    I_2&={\rm det}\,\msb{C}=J^2=g/G,\\
    I_3&={\rm tr}{\,\bs \kappa}=G^{\alpha\beta}\kappa_{\alpha\beta},\\
    I_4&={\rm det}{\,\bs \kappa}=\f{1}{2}\mu^{\alpha\beta}\mu^{\gamma\delta}\kappa_{\alpha\gamma}\kappa_{\beta\delta},\\
    I_5&={\rm tr}(\msb{C}\bs{\kappa})=C_{\alpha\beta}\kappa^{\alpha\beta}=C^{\alpha\beta}\kappa_{\alpha\beta},\\
    I_6&={\rm tr}(\msb{C}\bs{\kappa}\bs{\mu})=G_{\alpha\beta}C_{\gamma\delta}\kappa^{\alpha\gamma}\mu^{\beta\delta}=G_{\alpha\beta}\kappa_{\gamma\delta}C^{\alpha\gamma}\mu^{\beta\delta}.
    \end{split}
\label{e:invariants}
\end{align}

The choice of these basis invariants are unique to hemitropic material property. These invariants mentioned above were first listed by \citet{zheng1993two} and further, the two-dimensional Cayley-Hamilton theorem can be used to obtain the equivalence of these invariants to that provided by Zheng.
The expressions for the Cauchy interface stress $(\bs \sigma)$ and the moment $(\msb m)$ tensors are calculated using the relations
\begin{align}
    {\bs \sigma}=2\,\f{\del \gamma}{\del\msb C},\quad {\msb m}=\f{\del\gamma}{\del \bs \kappa}.
\end{align}
Using the expression for interface energy in terms of invariants, one can obtain the expressions for the components of interface stress tensor and the Eulerian bending moment tensor as
\begin{align}
\begin{split}
    \f{1}{2}J\sigma^{\alpha\beta}=\f{\del \gamma}{\del I_1}G^{\alpha\beta}+
    \f{\del \gamma}{\del I_2}\td{C}^{\alpha\beta}+\f{\del \gamma}{\del I_5}\kappa^{\alpha\beta}+\f{1}{2}\f{\del \gamma}{\del I_6}\left(D^{\alpha\beta}+D^{\beta\alpha}\right),\\
    Jm^{\alpha\beta}=\f{\del \gamma}{\del I_3}G^{\alpha\beta}+
    \f{\del \gamma}{\del I_4}\td{\kappa}^{\alpha\beta}+\f{\del \gamma}{\del I_5}C^{\alpha\beta}+\f{1}{2}\f{\del \gamma}{\del I_6}\left(E^{\alpha\beta}+E^{\beta\alpha}\right),
    \label{e:istr}
\end{split}
\end{align}
where
\begin{align}
    D^{\alpha\beta}=G_{\gamma\delta}\mu^{\alpha\gamma}\kappa^{\beta\delta},\quad
    E^{\alpha\beta}=G_{\gamma\delta}\mu^{\alpha\gamma}C^{\beta\delta}.\nn
\end{align}
Here, $\msb{\td{C}}$ and $\bs{\td{\kappa}}$ are the adjugate of $\msb{C}$ and $\bs{\kappa}$, respectively, with components given by
\begin{align}
    \td{C}^{\alpha\beta}&=({\rm tr}\,\msb{C})G^{\alpha\beta}-C^{\alpha\beta}, \\
    \td{\kappa}^{\alpha\beta}&=({\rm tr}\,\bs{\kappa})G^{\alpha\beta}-\kappa^{\alpha\beta}.
\end{align}
We assume that the bubble undergoes deformation from a FC such that the bulk deformation tensor $\msb{A}$ is compatible with the interface deformation tensor $\msb{a}$. From the balance of linear momentum, the general governing equation for the radial dynamics of the encapsulation and the outer liquid can be expressed in terms of bulk stress field $\msb T$ and the velocity field
$\bs{u}(r,t)$ as
\begin{align}
    \varrho\left[{\bs u}_{,t}+(\bs u\cdot\bs \nabla){\bs u}\right]=\bs \nabla\cdot\msb T,
\end{align}
where $\varrho$ is the density of the medium. Due to the existence of interface energy, the generalized Young-Laplace interface jump condition can be expressed in terms of interface stress and bending moment as (for detailed derivations, see \citet{steigmann, GAO201459}) 
\begin{align}
[\![\msb T]\!]\bs{\cdot n}=\left[(\sigma^{\alpha\beta}-m^{\lambda\alpha}q^{\beta}_{\;\;\lambda})\bs{g}_\beta+m^{\beta\alpha}_{\quad:\beta}\bs{n}\right]_{:\alpha},
\label{e:YL}
\end{align}
where the surface covariant derivative of a second order tensor $m^{\beta\alpha}_{\;\,\;\,:\beta}$ is given by the expression
\begin{align}
    m^{\beta\alpha}_{\;\,\;\,:\beta}=m^{\beta\alpha}_{\;\,\;\,,\beta}+m^{\delta\alpha}\,\Gamma^\beta_{\;\,\delta\beta}+m^{\beta\delta}\,\Gamma^\alpha_{\;\,\delta\beta},
\end{align}
and $[\![\msb T]\!]$ is the jump in the stress due to interface energy.
Similar to the calculations of $Q_{\alpha\beta}$ in \eqref{e:Qab}, using the relations \eqref{e:gdown},\eqref{e:gup} and \eqref{e:normal} we obtain the expression for the curvature tensor for the {\it inner} (minus sign) and {\it outer} (plus sign) interfaces as
\begin{align}
\begin{split}
    q_{\alpha\beta}&=-(\mp r(r_0,t)^{-1}g_{\alpha\beta})=-(\mp r_0^{-1}\lambda G_{\alpha\beta}),\\
    q^\alpha_{\;\;\beta}&=g^{\alpha\gamma}q_{\beta\gamma}=-(\mp r_0^{-1}\lambda^{-1} \delta^\alpha_{\;\;\beta}),
\end{split}
\end{align}
and the expressions for the components of the relative curvature tensor $\bs{\kappa}$ are given by
\begin{align}
\begin{split}
    \kappa_{\alpha\beta}&=\mp r_0^{-1}\lambda G_{\alpha\beta},\\
    \kappa^{\alpha\beta}&=\mp r_0^{-1}\lambda^3 g^{\alpha\beta}.
\end{split}
\end{align}
Therefore, the contra and co-variant components of the interface deformation tensor components can be written as
\begin{align}
\begin{split}
    C_{\alpha\beta}&=\lambda^2 G_{\alpha\beta},\\
    C^{\alpha\beta}&=G^{\alpha\gamma}G^{\beta\delta}C_{\gamma\delta}=\lambda^4g^{\alpha\beta}.  
\end{split}
\end{align}
The invariants in \eqref{e:invariants} are calculated as
\begin{align}
    I_1=2\,\lambda^2, \quad I_2=\lambda^4, \quad I_3=\mp2\,\lambda/r_0,
    \quad I_4=\lambda/r_0^2, \quad I_5=\mp2\,\lambda^3/r_0, \quad I_6=0.
\end{align}

For the undeformed interface configuration, the identity tensor (unit tensor) $\bs{1}$ is given by 
$$\bs{1}=G_{\alpha\beta}\,\bs{G}^\alpha\otimes\bs{G}^\beta.$$
Using identity tensor one can write $\bs{\rm C}=\lambda^2\bs{1}$ and the components of the adjugate of $\msb{C}$ are $\td{C}^{\alpha\beta}=C^{\alpha\beta}.$ From the relations in \eqref{e:istr}, components of interface stress and Eulerian bending moment tensors, respectively, can be simplified to yield the relations
\begin{align}
    \sigma^{\alpha\beta}&=\sigma g^{\alpha\beta},\quad
    m^{\alpha\beta}=m g^{\alpha\beta},
\end{align}
where 
\begin{align}
    \sigma&=\gamma_0+2\left(\gamma_1+\gamma_2\lambda^2\mp\f{\gamma_5\lambda}{r_0}\right),\\
    m&=\gamma_3\mp\f{\gamma_4\lambda}{r_0}+\gamma_5\lambda^2,
\end{align}
with `$-$' and `$+$' signs are supposed to be used for the {\it inner} and {\it outer} interfaces, respectively. Here, deformation independent interface tension is introduced as $\gamma_0$, and $\gamma_i=\del \gamma/\del I_i, i=1,2,$ are considered as interface material parameters. The generalized Young-Laplace equation \eqref{e:YL} for the radial dynamics can be expressed as
\begin{align}
    [\![\msb T]\!]\bs{\cdot n}=-(\mp 2 r_0^{-1}\lambda^{-1}\Sigma)\;\bs{n},
\end{align}
where $\Sigma=\sigma\mp r_0^{-1}\lambda^{-1}m$. The residual interface stress and bending moment tensors at static equilibrium  are calculated by the substitution $\lambda=1$, which yields $\sigma^{*\alpha\beta}=\sigma^*g^{\alpha\beta}$ and $m^{*\alpha\beta}=m^*g^{\alpha\beta}$, where $\sigma^*$ and $m^*$ are given by
\begin{align}
\sigma^*&=\gamma_0+2\left(\gamma_1+\gamma_2\mp\f{\gamma_5}{r_0}\right),\\ m^*&=\gamma_3+\gamma_5\mp\f{\gamma_4}{r_0}.
\end{align}
\textcolor{black}{Here, the assumption of $\gamma_i=\del \gamma/\del I_i$ as constant gives a simple linear structure to the interface energy density function with a linear combination of the basis invariants defined in \eqref{e:invariants}. However, in a more general case, the interface energy can be a nonlinear function of these basis invariants. Therefore, there is always scope to improve/modify the mathematical structure of the Young-Laplace equation and the residual stress field}.

Comparing the residual stress $\sigma^*$ at the interface with two leading order terms of well known Tolman's formula \citep{tolman1949effect} $\sigma_s\sim\sigma_\infty(1-2\delta_\infty/r_0)$ for a curved interface indicates that $\gamma_5$ has the role of Tolman's length scale $\delta_\infty$. In $\Sigma$, $\gamma_3$ appears to have the similar interpretation as $\gamma_5$ for $\lambda=1$, however, it is important to identify the distinct effects of $\gamma_3$ and $\gamma_5$, on the interface tension as will be shown in equation~\eqref{e:sigma12}.

As mentioned earlier, the present model naturally introduces residual stress in the bulk due to the expansion/shrinkage of the bubble after preparation, which was also observed in the experiments \citep{DEJONG2007}. This also indicates the important feature of equivalent NC bubbles with the same radii and internal gas pressure at NC, which can show completely different mechanical behavior. This behavior is strongly connected to the process of attaining NC, which in-turn is related to the interface energy induced residual stress. On the other hand, one would also expect that different size bubbles made of the same encapsulation should demonstrate the behavior based on same material constants. Here, we demonstrate these features by considering the above interface energy and corresponding interface parameters in the mathematical model of encapsulated bubble dynamics.

\subsection{Constitutive relation for the shell and the fluid}
\label{sec:cr}
We analyze the EB dynamics by assuming that the encapsulation/shell and the fluid outside/surrounding the bubble are homogeneous, isotropic, and incompressible materials. Further, the shell is assumed to be viscoelastic. A Kelvin-Voigt type constitutive model with elastic part described by Mooney-Rivlin (MR) material model with elastic constants $C_1$ and $C_2$, and viscous part by the Newton's law of viscosity is considered. 
In order to capture the material nonlinearity of the encapsulation, we have chosen this basic nonlinear material model. The ratio $C_2/C_1$ in the MR material model is usually interpreted as a strain-softening/hardening parameter depending on its value.
For an incompressible bulk MR material, the strain energy density $W=C_1(\td{I}_1-3)+C_2(\td{I}_2-3)$, where $\td{I}_1,\;\td{I}_2,\;\td{I}_3$ are the three invariants of $\msb B$ given by
\begin{align}
\begin{split}
    \td I_1&={\rm tr}{\,\msb B}=2\Lambda^2+\Lambda^{-4},\\
    \td I_2&={\rm tr}{\,\msb B^2}=2\Lambda^{-2}+\Lambda^4,\\
    \td I_3&={\rm det}\,\msb B=1,
\end{split}
\end{align}
where $\Lambda=r/r_e$ is the principal stretch ratio in both $\theta^1$ and $\theta^2$ directions. Since the bulk is assumed to be incompressible, ${\td I}_3=1$ gives 
\begin{align}
    \f{\partial{r}}{\partial r_e}=\left(\f{r_e}{r}\right)^2=\Lambda^{-2}.
    \label{e:incomp}
\end{align}
Using superscripts S and L for the shell and the liquid parameters, respectively, the Cauchy stress tensor for the shell is given by
\begin{align}
    \msb T^{\rm S}=-p^{\rm S}\msb{I}+2(C_1+\td{I}_1C_2)\msb{B}-2C_2\msb{B}^2+\eta^{\rm S}\left[\bs\nabla\bs u^{\rm S}+(\bs\nabla\bs{u}^{\rm S})^{\rm T}\right],
    \label{e:TS}
\end{align}
where, $\eta^{\rm S}$ is the shell viscosity coefficient, $\msb{I}$ is the identity tensor and $p^{\rm S}$ is commonly referred to as arbitrary hydrostatic pressure due to incompressibility. The liquid viscous stress tensor with viscosity coefficient $\eta^{\rm L}$ is given by the constitutive equation 
\begin{align}
    \msb T^{\rm L}=-p^{\rm L}{\msb I}+\eta^{\rm L}\left[\bs\nabla\bs u^{\rm L}+(\bs\nabla\bs{u}^{\rm L})^{\rm T}\right].
    \label{e:TL}
\end{align}

\subsection{Governing equation}
\label{sec:ge}
For the radial dynamics of the shell exerted by an ultrasound field and the surrounding liquid, the governing equation in spherical polar coordinates with $(r,\phi,\theta)$ as principal coordinate directions can be reduced to radial equation
\begin{align}
    \varrho\left(\f{\del u}{\del t}+{u}\f{\del u}{\del r}\right)=\f{\del T_{rr}}{\del r}+\f{2 T_{rr}-T_{\phi\phi}-T_{\theta\theta}}{r}.
\label{e:rad}
\end{align}
The stress jump conditions at the inner and outer interfaces are 
\begin{align}
\begin{split}
    T_{rr}^{\rm S}\big\lvert_{r=R_1}+p_{g}&=\f{2\sigma_1}{R_1},\\
    \left[T_{rr}^{\rm S}-T_{rr}^{\rm L}\right]_{r=R_2}&=-\f{2\sigma_2}{R_2},
    \label{e:jump}
\end{split}
\end{align}
where $\sigma_1$ and $\sigma_2$ are the interface tension parameters given by 
\begin{align}
\begin{split}
    \sigma_1&=\gamma_{10}+2\left(\gamma_{11}+\gamma_{12}\f{R_1^2}{R_{10}^2}-\gamma_{15}\f{R_1}{R_{10}^2}\right)-\f{1}{R_1}\left(\gamma_{13}-\gamma_{14}\f{R_1}{R_{10}^2}+\gamma_{15}\f{R_1^2}{R_{10}^2}\right),\\
    \sigma_2&=\gamma_{20}+2\left(\gamma_{21}+\gamma_{22}\f{R_2^2}{R_{20}^2}+\gamma_{25}\f{R_2}{R_{20}^2}\right)+\f{1}{R_2}\left(\gamma_{23}+\gamma_{24}\f{R_2}{R_{20}^2}+\gamma_{25}\f{R_2^2}{R_{20}^2}\right),
\end{split}
\label{e:sigma12}
\end{align}
with $\gamma_{i(\cdot)}, i=1,2$ correspond to inner and outer interface, respectively. Unlike existing shell models, interface tensions are considered on either side of the bubble shell and dependent on the interface radius.

Substituting $R_1=R_{10}$ and $R_2=R_{20}$ in the stress jump conditions in (\ref{e:jump}), one can rewrite the interface conditions at the natural configuration of bubble as
\begin{align}
    T_{rr}^{\rm S}\bigg\lvert_{R_1=R_{10}}+p_{g_0}&=\f{2}{R_{10}}\left[\gamma_{10}+2\left(\gamma_{11}+\gamma_{12}-\f{\gamma_{15}}{R_{10}}\right)-\f{1}{R_{10}}\left(\gamma_{13}-\f{\gamma_{14}}{R_{10}}+\gamma_{15}\right)\right],\\
    T_{rr}^{\rm S}\bigg\lvert_{R_2=R_{20}}-T_{rr}^{\rm
    L}\bigg\lvert_{R_2=R_{20}}&=\f{-2}{R_{20}}\left[\gamma_{20}+2\left(\gamma_{21}+\gamma_{22}+\f{\gamma_{25}}{R_{20}}\right)+\f{1}{R_{20}}\left(\gamma_{23}+\f{\gamma_{24}}{R_{20}}+\gamma_{25}\right)\right].
\label{e:prestress}
\end{align}
To obtain the governing equation, integrating equation~(\ref{e:rad}) with respect to radial coordinate $r$ for the shell and the liquid regions separately, we have
\begin{align}
    &\int_{R_1}^{R_2} \varrho^{\rm S}\left(\f{\del u^{\rm S}}{\del t}+{u^{\rm S}}\f{\del u^{\rm S}}{\del r}\right) {\rm d}r+\int_{R_2}^{\infty} \varrho^{\rm L}\left(\f{\del u^{\rm L}}{\del t}+{u^{\rm L}}\f{\del u^{\rm L}}{\del r}\right) {\rm d}r= \nn \\ 
    &\int_{R_1}^{R_2} \left(\f{\del T^{\rm S}_{rr}}{\del r}+\f{2T_{rr}^{\rm S}-T^{\rm S}_{\phi\phi}-T^{\rm S}_{\theta\theta}}{r}\right) {\rm d}r+\int_{R_2}^\infty \left(\f{\del T^{\rm L}_{rr}}{\del r}+\f{2T^{\rm L}_{rr}-T^{\rm L}_{\phi\phi}-T^{\rm L}_{\theta\theta}}{r}\right) {\rm d}r,
\label{e:inteq}
\end{align}
where the superscripts S and L represent the physical terms corresponding to shell and liquid, respectively. 
From \eqref{e:TS}, we calculate the expressions for the components of the Cauchy stress tensor for the shell as
\begin{align}
\begin{split}
    T^{\rm S}_{rr}&=-p^{\rm S}+2\lambda^{-4}C_1+4\lambda^{-2}C_2+2\eta^{\rm S}\left(\f{\del u^{\rm S}}{\del r}\right),\\
    T^{\rm S}_{\phi\phi}&=T^{\rm S}_{\theta\theta}=-p^{\rm S}+2\lambda^{2}C_1+2(\lambda^4+\lambda^{-2})C_2+2\eta^{\rm S}\left(\f{u^{\rm S}}{r}\right),\\
    T^{\rm S}_{r\phi}&=T^{\rm S}_{\phi\theta}=T^{\rm S}_{\theta r}=0.
    \label{e:TScomp}
\end{split}
\end{align}
Similarly from \eqref{e:TL}, we calculate the components of liquid viscous stress tensor as
\begin{align}
\begin{split}
    T^{\rm L}_{rr}&=-p^{\rm L}+2\eta^{\rm L}\left(\f{\del u^{\rm L}}{\del r}\right),\\
    T^{\rm L}_{\phi\phi}&=T^{\rm L}_{\theta\theta}=-p^{\rm L}+2\eta^{\rm L}\left(\f{u^{\rm L}}{r}\right),\\
    T^{\rm L}_{r\phi}&=T^{\rm L}_{\phi\theta}=T^{\rm L}_{\theta r}=0.
    \label{e:TLcomp}
\end{split}
\end{align}
Using the above expressions for the stress components normal to the three principal coordinate planes and the expression for velocity field in Appendix \ref{app:A}, the governing equation for the radial dynamics of a bubble exerted by an ultrasound field is given by 
\begin{align}
\varrho^{\rm S}\left[R_1\ddot{R}_1+\f{3}{2}\dot{R}_1^2\right]-\left(\varrho^{\rm S}-\varrho^{\rm L}\right)\left[R_2\ddot{R}_2+\f{3}{2}\dot{R}_2^2\right]&=p_g-p_0-p_d+p_e-\f{2\sigma_1}{R_1}-\f{2\sigma_2}{R_2}\nn\\
&-4\eta^{\rm L}\f{\dot{R}_2}{R_2}-4\eta^{\rm S}\left(\f{\dot{R}_1}{R_1}-\f{\dot{R}_2}{R_2}\right),
\label{e:neq}
\end{align}
where $p_g$ is the gas pressure, $p_0$ is the ambient pressure, $p_d=p_a\sin(2\pi f t)$ is the external acoustic field with pressure $p_a$ and frequency $f$, and  $p_e$ is the elastic restoring force (see Appendix \ref{app:B}) given by
\begin{align}
    p_e=&-4\left[C_1\left(\f{1}{4\Lambda_2^4}+\f{1}{\Lambda_2}\right)+C_2\left(\f{1}{2\Lambda_2^2}-\Lambda_2\right)\right]+4\left[C_1\left(\f{1}{4\Lambda_1^4}+\f{1}{\Lambda_1}\right)+C_2\left(\f{1}{2\Lambda_1^2}-\Lambda_1\right)\right].
\label{e:elarf}
\end{align}
Here, $\Lambda_1=R_1/R_{e_1}$ and $\Lambda_2=R_2/R_{e_2}$ are the stretch ratios. The interface tension parameters $(\sigma_1,\sigma_2)$ and the elastic restoring force $(p_e)$ are the additional terms which makes the governing equation (\ref{e:neq}) different from that of the Church's model \citep{Church1995}. When the bubble is in its natural configuration the time dependent terms in the governing equation vanish and the static equilibrium equation is given by
\begin{align}
    p_{g_0}=p_0&+\f{2}{R_{10}}\left[\gamma_{10}+2\left(\gamma_{11}+\gamma_{12}-\f{\gamma_{15}}{R_{10}}\right)-\f{1}{R_{10}}\left(\gamma_{13}-\f{\gamma_{14}}{R_{10}}+\gamma_{15}\right)\right]\nn\\
    &+\f{2}{R_{20}}\left[\gamma_{20}+2\left(\gamma_{21}+\gamma_{22}+\f{\gamma_{25}}{R_{20}}\right)+\f{1}{R_{20}}\left(\gamma_{23}+\f{\gamma_{24}}{R_{20}}+\gamma_{25}\right)\right]\nn\\
    &+4\left[C_1\left(\f{1}{4\Lambda_{20}^4}+\f{1}{\Lambda_{20}}\right)+C_2\left(\f{1}{2\Lambda_{20}^2}-\Lambda_{20}\right)\right]\nn\\
    &-4\left[C_1\left(\f{1}{4\Lambda_{10}^4}+\f{1}{\Lambda_{10}}\right)+C_2\left(\f{1}{2\Lambda_{10}^2}-\Lambda_{10}\right)\right].
\label{e:pg0}
\end{align}
Let $p_{g_e}$ be the uniform gas pressure inside the bubble in the fictitious configuration. Assuming that the gas inside the bubble is undergoing expansion/compression with polytropic expansion index $k=1.4$, the relation between $p_g,p_{g_0}$ and $p_{g_e}$ can be expressed as
\begin{align}
    p_g=p_{g_0}\left(\f{R_{10}}{R_1}\right)^{3k}=p_{g_e}\left(\f{R_{e_1}}{R_1}\right)^{3k}.
\end{align}
With the inner and outer radii of the bubble at natural configuration being known, the bubble radii at fictitious configuration can be found by solving the nonlinear algebraic equation~(\ref{e:pg0}).

As we are interested in bubbles of radius $\textit{O}(10^{-6})\,$m and thickness $\textit{O}(10^{-9})\,$m, inspecting the terms in equation~\eqref{e:sigma12}, it is reasonable to assume the order of parameters $\gamma_{10},\gamma_{20},\gamma_{11},\gamma_{21}$ as $\textit{O}(1)\,$N/m, $\gamma_{13},\gamma_{23},\gamma_{15},\gamma_{25}$ as $\textit{O}(10^{-6}\,$)N and $\gamma_{14},\gamma_{24}$ as $\textit{O}(10^{-12})\,$N\,m. Primarily this assumption ensures that the effective interface tension parameters $\sigma_1$ and $\sigma_2$ possess reasonable values of $\textit{O}(1)$N/m. 
Also, it is worth mentioning that the order of these interface parameters are chosen such that it will not lead to large deviations in the bubble radii while undergoing deformation from the fictitious configuration to the natural configuration.
In the rest of the paper, $\varrho^{\rm S}=1100\;\rm{kg/m^3}$, $\varrho^{\rm L}=1000\;\rm{kg/m^3}$, $\gamma_{10}=0.04\;$N/m, $\gamma_{20}=0.005\;$N/m and a numerical value of $\gamma_{ij}$ by default carries these orders apart from any other explicitly mentioned factors.

 In equation~\eqref{e:sigma12}, $\gamma_{i2}$ term retains the effect of relative interface area. However, in the present model, the dynamic interface tension cannot become zero due to the change in the interface area, unlike the other models \citep{Paul2010}. On the other hand, both $\gamma_{13}$ and $\gamma_{15}$ can make the dynamic interface tension negative at the inner interface, which is purely an effect of interface bending rigidity. These relations also show distinct effects of $\gamma_3$ and $\gamma_5$ terms when the interface starts deforming.  It is interesting to observe that contribution of $\gamma_5$ terms in equation~\eqref{e:neq} is only through static equilibrium pressure and depends only on the initial size of the bubble.  On the other hand, the dynamic interface tension has terms proportional to the interface radius and curvature through $\gamma_5$ and $\gamma_3$, respectively, which shows distinct effects. 

Here, we study the influence of three important interface parameters $\gamma_2,\gamma_3$, and $\gamma_5$ on EB's natural configuration and radial dynamics. These three interface parameters are related to the deformation dependent interface area, curvature, and the coupling between the two. To understand the influence of these aspects further, we have constructed a multistart optimization problem by defining a compression/expansion ratio to estimate the interface and material parameters.

\section{Constrained optimization}
\label{sec:co}
Let $R_2^{\rm max}$ and $R_2^{\rm min}$ be the maximum and minimum outer radii, respectively, of the steady state response of a bubble at particular excitation pressure and frequency. The compression/expansion ratio can be defined as $$\zeta =\frac{R_2^{\rm max}-R_{20}}{R_{20}-R_2^{\rm min}}.$$ Here, $\zeta\ll1 (\gg1)$ represents the compression (expansion) dominant behavior of a bubble. 

Based on typical compression dominant radial dynamics observed in the experiments at a specific range of excitation pressure and frequency \citep{Doinikov2011}, suitable parameters are estimated by setting up an minimization problem with parameters $C_1,C_2,\eta^{\rm L},\eta^{\rm S},f,$ and $\gamma_{ij}, i=1,2, j=1$ to $5,$ as
\begin{align}
\begin{split}
    \text{Minimize:\quad}&  \zeta \\
    \text{Constraints:\quad} & \zeta \leq \zeta_0,\;0.5 \leq \zeta_0 \leq 1,\\
    & \td R_{e_1}\geq 0.75,\td R_{e_2} \leq 0.95\;\text{(Case \hypertarget{e:caseI}{\hyperlink{tb:caseI}{I }}: expanding
    bubble)},\\
    & \td R_{e_1}\geq 1.05,\td R_{e_2} \leq 1.25\;\text{(Case \hypertarget{e:caseII}{\hyperlink{tb:caseII}{II}}: shrinking bubble)}
\end{split}
\end{align}
Here, $\zeta_0$ is the compression ratio obtained for a bubble attaining equivalent NC without interface energy and $\td R_{e_i}=R_{e_i}/R_{10}$ is the non-dimensional inner $(i=1)$ and outer $(i=2)$ radii of the bubble in FC. We have solved the optimization problem using Matlab\textsuperscript{\textregistered} R2018a (9.4.0) multistart optimization.

The above optimization problem is solved with proper nonlinear constraints for a bubble of $2\,\mu$m radius with $20\,$nm thickness. The interface and material parameters are estimated for dominant compression behavior, which will be discussed in the results section. Here, the shell thickness and other shell properties are chosen, referring to the agent bubble \textit{Optison}\texttrademark.

To fit the experimental data \citep{vanderMeer2007,Doinikov2009,DEJONG2007,Tu2009} for thin lipid monolayer EBs, the value of $\zeta$ is first identified from the experimental time-series data. Using this value as a reference in the optimization problem, the constraints for $\zeta$ are modified close to experimental observations to estimate the interface and material parameters. In the simulations, the only information from the experiments used to fit the radius-time curves is the experimental $\zeta$ value, $R_2^{\rm min}$, excitation pressure, and frequency.

\section{Results and discussion}
\label{sec:results}

\subsection{Analysis of various bubble configurations}
A bubble can attain its NC by shrinkage or expansion from a stress free configuration as shown in figure~\ref{f:bubble}. Let $\Lambda_{10}=R_{10}/R_{e1}$ is the pre-stretch at the inner interface with $\Lambda_{10} < 1 (>1)$ for the bubble attaining equilibrium NC by shrinking (expanding). For a given set of optimized interface parameters (IP) and material parameters (MP) with assumed equilibrium gas pressure, $p_{g_0}$, one can obtain the equation for static (natural) equilibrium from \eqref{e:neq}, which has a unique solution for FC radii. Change in any of the IP data possibly leads to different FC, compression/expansion ratio, and pre-stress in the bulk. However, both these data represent equivalent NC if we assume that the amount of gas inside the bubble is the same. Therefore, we compare such configurations to understand the influence of interface parameters on the bubble dynamics through compression/expansion ratio and pre-stress.

\begin{table}
\begin{center}
\begin{tabular}{ccccccccccc}
~ & \multicolumn{10}{c}{{\bf Interface parameters} $\gamma_{ij}\left(\times10^{-2}\right)$}\\
Case \quad & $\gamma_{11}$ \quad & $\gamma_{21}$ \quad & $\gamma_{12}$ \quad & $\gamma_{22}$ \quad & $\gamma_{13}$ \quad & $\gamma_{23}$ \quad &  $\gamma_{14}$ \quad & $\gamma_{24}$ \quad & $\gamma_{15}$ \quad & $\gamma_{25}$\\
\hypertarget{tb:caseI}{\hyperlink{e:caseI}{I}} \quad & 1.00 \quad & 1 \quad & 4.19 \quad & 6.19 \quad & 7.89 \quad & 1 \quad & 1 \quad & 1 \quad & 8.00 \quad & 1 \\
\hypertarget{tb:caseII}{\hyperlink{e:caseII}{II}} \quad & 1.01 \quad & 1 \quad & 6.86 \quad & 8.77 \quad & 8.00 \quad & 1 \quad & 1 \quad & 1 \quad & 7.99 \quad & 1 \\
\end{tabular}
\caption{Optimized interface parameters ({IP}) for a bubble with $R_{10}=2\;\mu$m, thickness $h=20\;$nm, natural configuration pressure $p_{g_0}=0.3\,$MPa and excitation pressure $p_a=0.1\,$MPa. Case I and II correspond to a bubble undergoing expansion and shrinkage, respectively, from fictitious to natural configuration.}
\label{tb:1}
\end{center}
\end{table}

\begin{table}
\begin{center}
\begin{tabular}{cccccccc}
~ & \multicolumn{7}{c}{\bf Bulk material parameters}\\
Case \quad & $C_1$ \quad & $C_2$ \quad & $\eta^{\rm L}$ \quad & $\eta^{\rm S}$ \quad & $f$ \quad & $\zeta$ \quad & $\zeta_0$ \\
\hypertarget{tb:caseI}{\hyperlink{e:caseI}{I}} \quad & 1.42 \quad & 1.39 \quad & 0.5 \quad & 0.01 \quad & 1.96 \quad & 0.57 \quad & 0.66 \\
\hypertarget{tb:caseII}{\hyperlink{e:caseII}{II}} \quad & 1.40 \quad & 0.67 \quad & 0.5 \quad & 0.01 \quad & 2.1~ \quad & 0.57 \quad & 0.83 \\
\end{tabular}
\caption{Optimized material parameters ({MP}) for a bubble with $R_{10}=2\;\mu$m, thickness $h=20\;$nm, natural configuration pressure $p_{g_0}=0.3\,$MPa and excitation pressure $p_a=0.1\,$MPa are shell elastic constants $\left(C_1,C_2\right)\;$MPa, viscosity of shell $\left(\eta^{\rm S}\right)$ Pa-s and liquid $\left(\eta^{\rm L}\right)\;$mPa-s, frequency $\left(f\right)\;$MHz, compression ratios $\zeta$ and $\zeta_0$ . Case I and II correspond to a bubble undergoing expansion and shrinkage, respectively, from fictitious to natural configuration.}
\label{tb:2}
\end{center}
\end{table}

\begin{figure}
\includegraphics[width=0.485\linewidth]{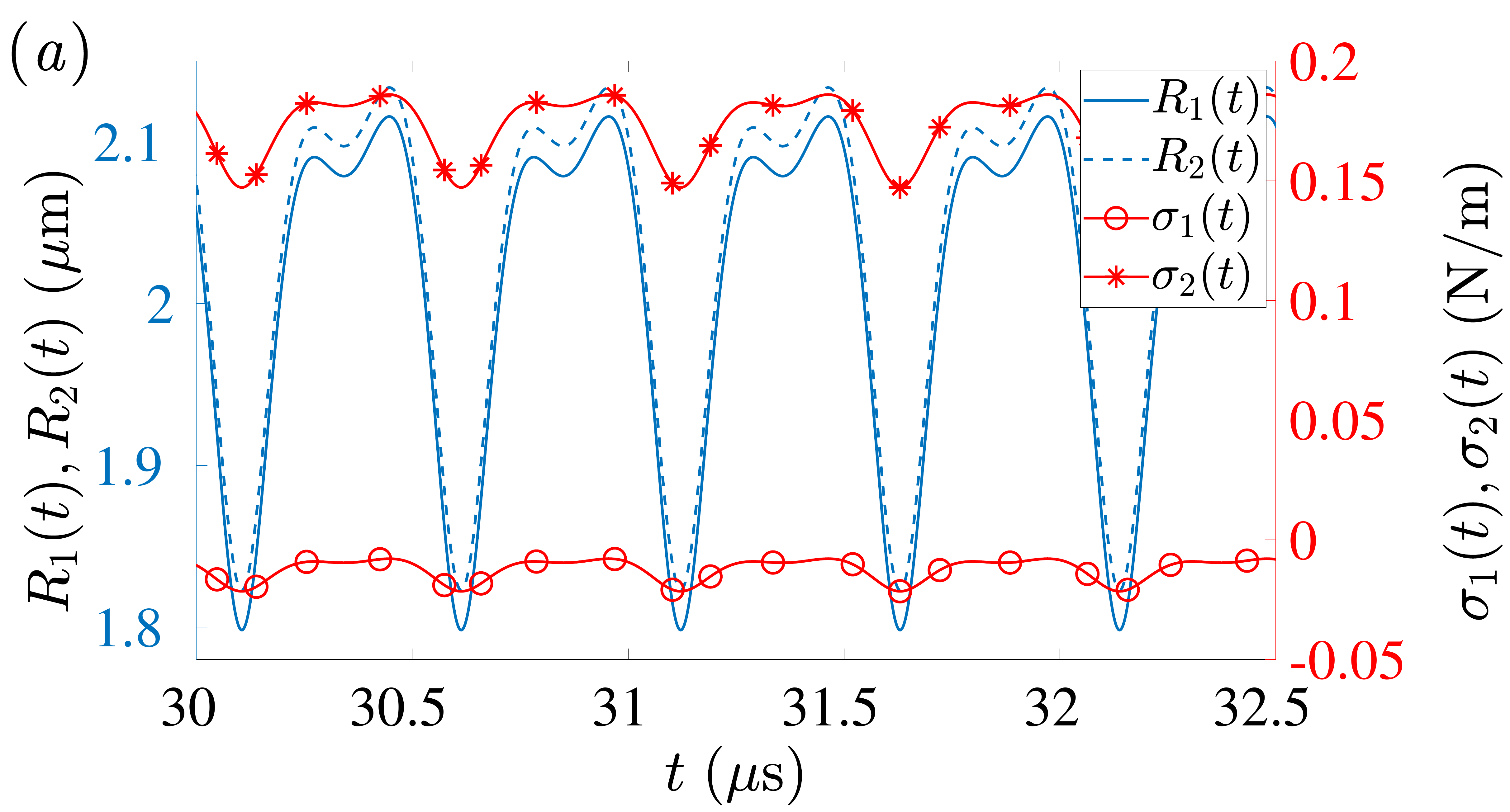}
\includegraphics[width=0.485\linewidth]{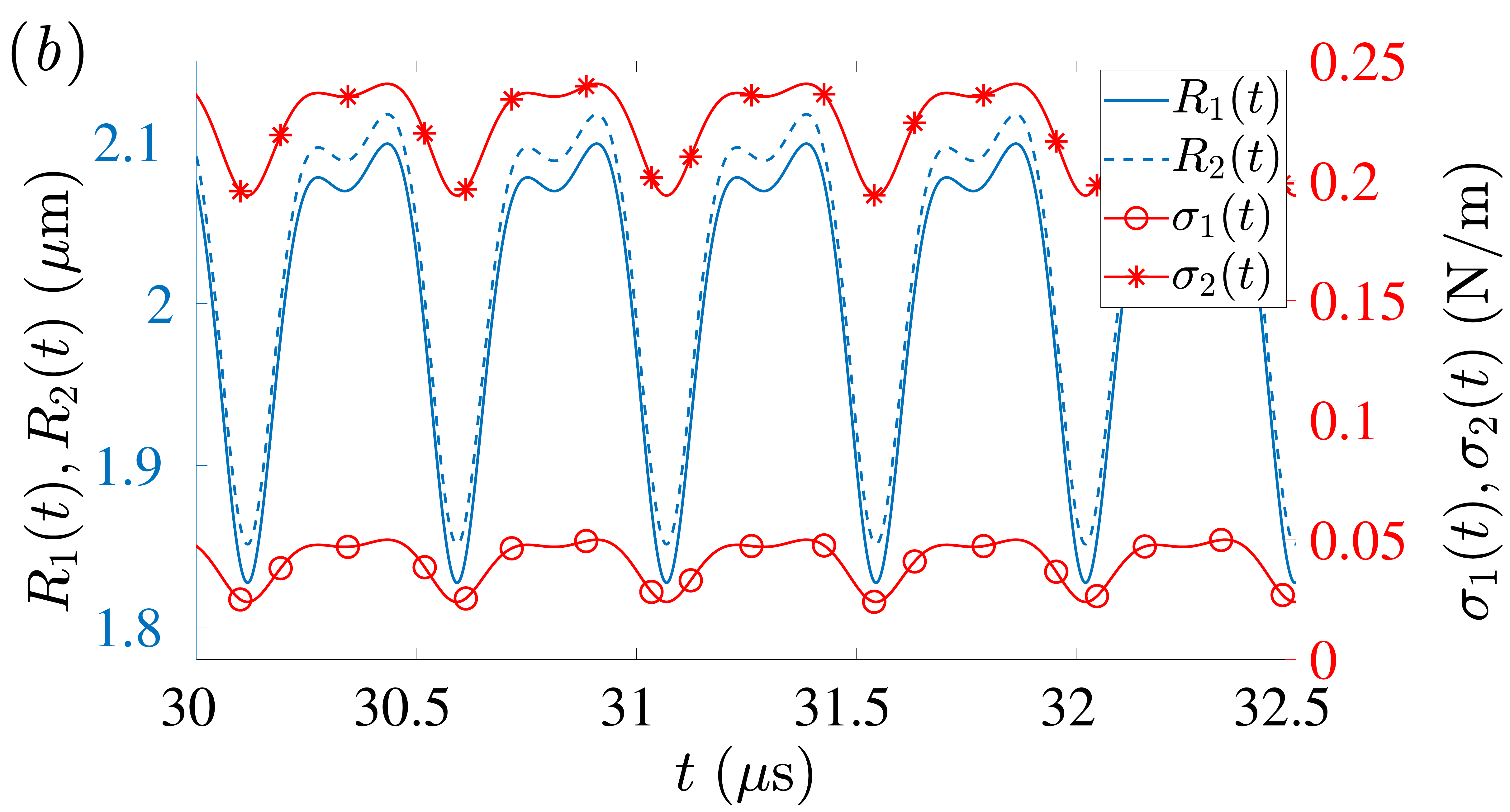}
\caption{Variation of bubble radii, $R_{i}(t)$, and interface tension, $\sigma_{i}(t)$, with time for the optimized values of interface (table~\ref{tb:1}) and material parameters (table~\ref{tb:2}) at $p_a=0.1\;$MPa for (\textit{a}) Case I and (\textit{b}) Case II.}
\label{f:2}
\end{figure}

Tables~\ref{tb:1} and \ref{tb:2} contain representative data of interface and material parameters obtained from the optimization for compression dominant behavior. It is interesting to note that interface energy induced bulk residual stress has an important role in the compression dominant behavior. To identify the dominant interface parameters of table~\ref{tb:1} data, bubbles with equivalent NC obtained from different FCs are compared in table~\ref{tb:3} by removing the contribution of specific interface parameters in the model. Data in table~\ref{tb:3} suggests that $\gamma_{i5}$ and $\gamma_{i2}$ $(i=1,2)$ have significant contributions for dominant compression. For instance, in case II, setting $\gamma_{i5}=0$ requires an increase in FC radii with $\zeta=1.24$, which is an expansion dominant behavior. On the other hand, setting $\gamma_{i2}=0$ shows a decrease in FC radii but still an increase in the value of $\zeta$. In the absence of interface energy, we have found an expanding FC in both the cases with $\zeta_0=0.66$. However, we exclude such cases because modeling of such interfaces is incomplete without interface energy. This discussion shows that the nature of interface energy induced pre-stress in the bulk influences the compression/expansion ratio ($\zeta$) and is highly dependent on the interface parameters. Similar dominance of $\gamma_{i5}$ and $\gamma_{i2}$ are also found for a bubble attaining NC from a FC (data not presented here). Another important feature observed in figure~\ref{f:2} for the above data is the negative and positive dynamic inner interface tensions for case I and case II, respectively, as expected from equation~\eqref{e:sigma12}. In this case, physically possible negative dynamic interface tension can be interpreted in terms of interface bending rigidity.

\begin{table}
\begin{center}
\begin{tabular}{ccccccc}
Case \quad \quad & Parameters \quad \quad & with all IP \quad \quad & $\gamma_{i2}=0$ \quad \quad & $\gamma_{i3}=0$ \quad \quad & $\gamma_{i4}=0$ \quad \quad & $\gamma_{i5}=0$ \\
\multirow{3}{*}{I} \quad \quad & $\zeta$ \quad \quad & 0.576 \quad \quad & 0.678 \quad \quad & 0.578 \quad \quad & 0.576 \quad \quad & 0.781                     \\
                   & $R_{e_1}$ \quad \quad & 1.877 \quad \quad & 1.458 \quad \quad & 1.977 \quad \quad & 1.862 \quad \quad & 2.195  \\
                   & $R_{e_2}$ \quad \quad & 1.900 \quad \quad & 1.495 \quad \quad & 1.997 \quad \quad & 1.886 \quad \quad & 2.212  \\
\multirow{3}{*}{II} \quad \quad & $\zeta$ \quad \quad & 0.575 \quad \quad & 0.831 \quad \quad & 0.588 \quad \quad & 0.576 \quad \quad & 1.249 \\
                    & $R_{e_1}$ \quad \quad & 2.250 \quad \quad & 1.303 \quad \quad & 2.383 \quad \quad & 2.231 \quad \quad & 2.620 \\
                    & $R_{e_2}$ \quad \quad & 2.266 \quad \quad & 1.349 \quad \quad & 2.397 \quad \quad & 2.247 \quad \quad & 2.632 \\ 
\end{tabular}
\caption{Values of compression ratio ($\zeta$), inner ($R_{e_1}$) and outer ($R_{e_2}$) radii ($\mu$m) of the bubble in fictitious configuration with the absence of specific IP $(\gamma_{ij},i=1,2$ and $j=2\;\rm{to}\;5)$ from table~\ref{tb:1}.}
\label{tb:3}
\end{center}
\end{table}

\subsection{Comparison with experimental data} 
To fit the experimental radius-time curves of EB, natural configuration gas pressure, $p_{g_0}$, and bubble thickness, $h$, are two important quantities apart from all the interface and material parameters. Their roles become significantly important as they provide an idea about how the bubble shell was produced/fabricated, as identified/discussed by \citet{doi:10.1121/10.0003500}.
For $1.7\,\mu$m radius bubble we modify the optimization problem with additional constraints $1.45<R_{2}^{\rm min}<1.5$ and optimization parameter, $p_{g_0}$. Referring to \citet{vanderMeer2007}, the thickness of the EB is considered to be $4\,$nm and $\eta^{\rm L}$ is taken as $2$ mPa-s. On a similar note, for $1.4\,\mu$m and $1\,\mu$m radius bubble, we modify the optimization problem with additional constraints $0.85<R_{2}^{\rm min}<0.9$ and $0.9<R_{2}^{\rm min}<0.95$, respectively, and optimization parameters, $p_{g_0}$ and $h$. Additionally in these simulations we include a constraint for shear modulus $(\mu)$ such that $15\leq \mu\,(\rm MPa)\leq 17$ (referring to the agent bubble \textit{Optison}\texttrademark). The optimized interface and material parameters thus obtained are listed in tables~\ref{tb:4} and \ref{tb:5}, respectively. We have fitted the experimental radius-time data for $1.7\,\mu$m \citep{vanderMeer2007}, $1.4\,\mu$m \citep{Doinikov2009} and $1\,\mu$m \citep{DEJONG2007} radius bubble with the present model using these optimized interface and material parameters.

\begin{figure}
\includegraphics[width=0.495\linewidth,trim=0in 0in 1.5in 0in,clip=true]{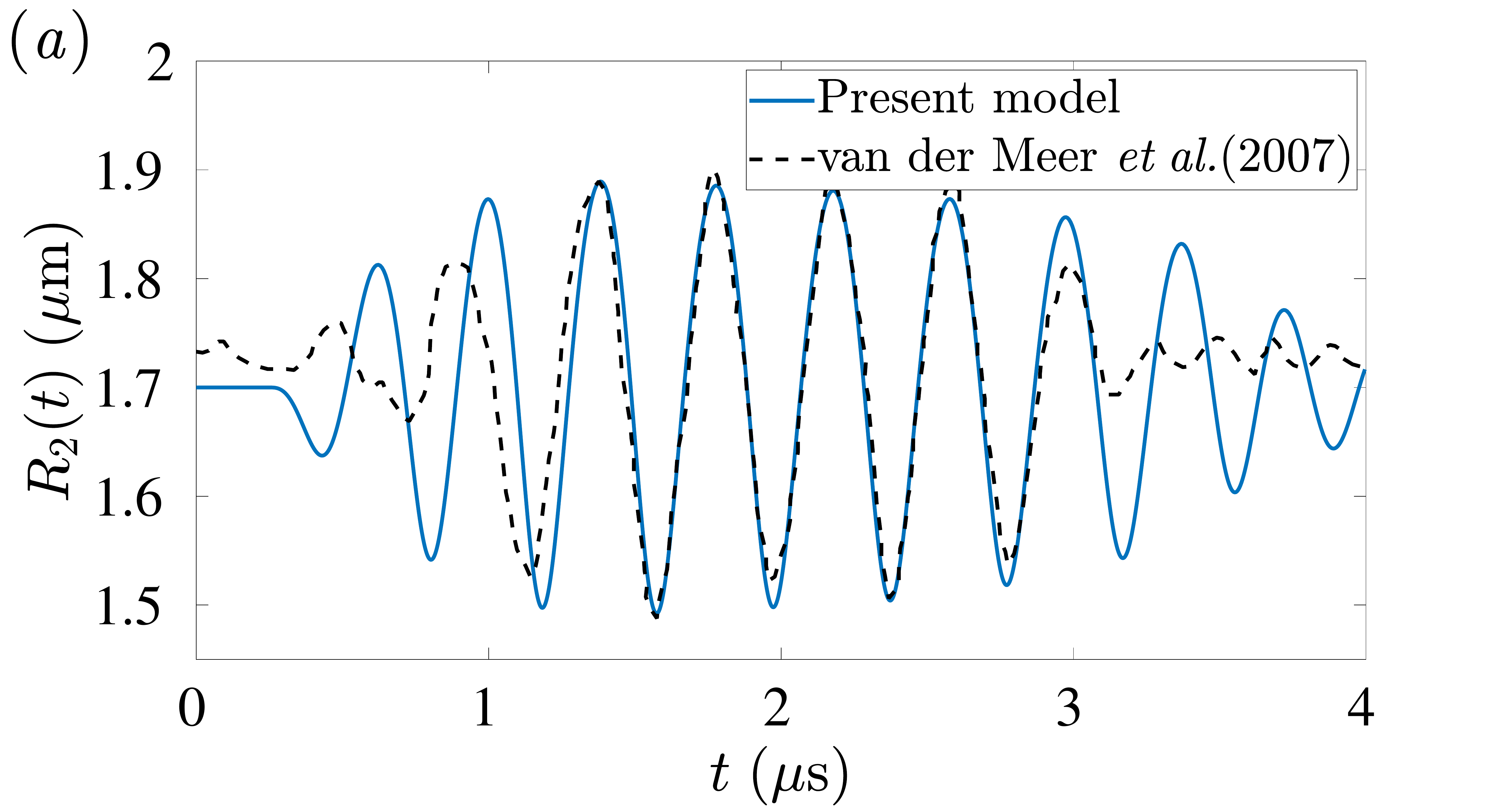}
\includegraphics[width=0.495\linewidth,trim=0in 0in 1.5in 0in,clip=true]{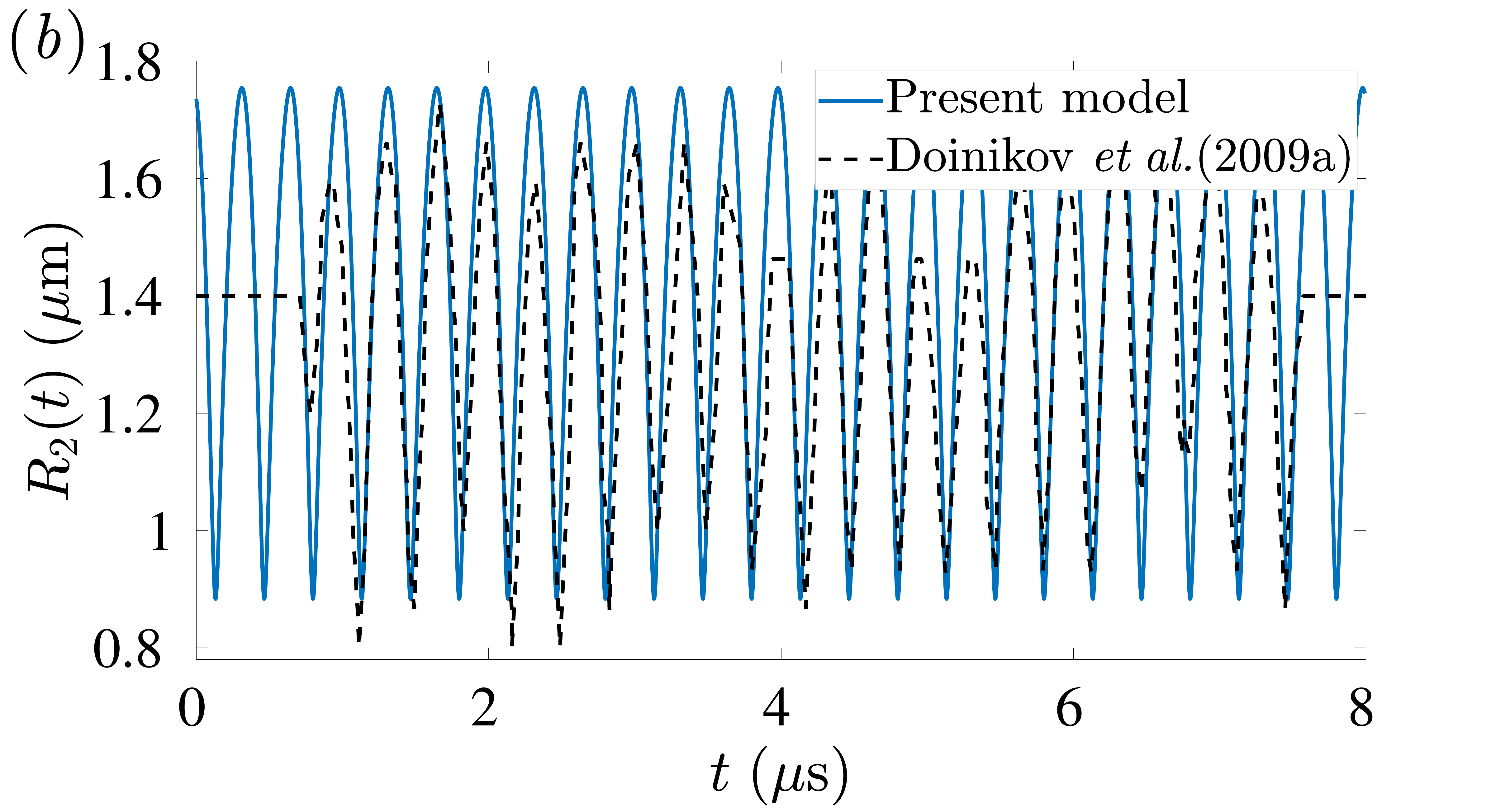}\\
\hspace*{0.25\linewidth}\includegraphics[width=0.5\linewidth,trim=0in 0in 1.5in 0in,clip=true]{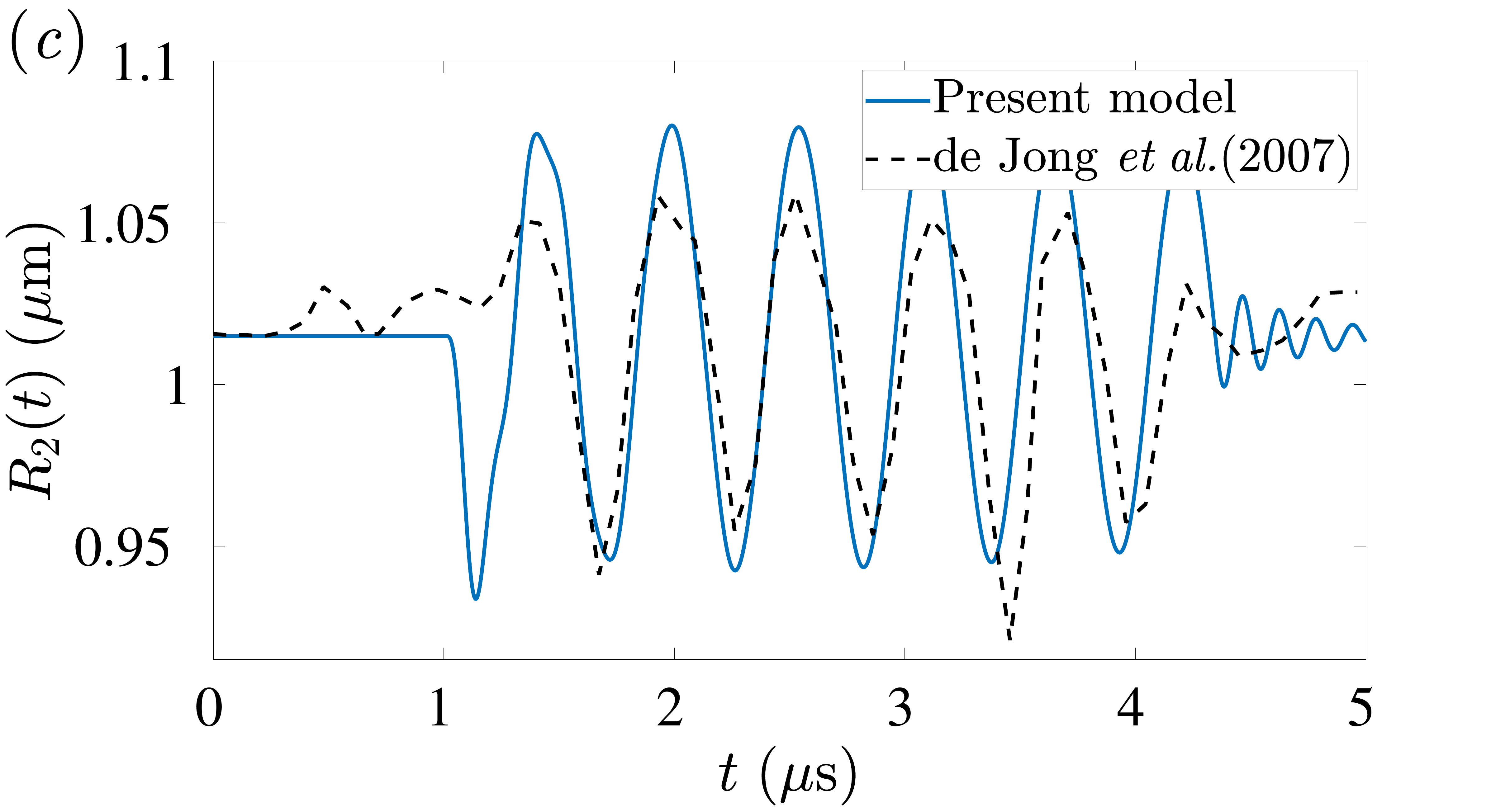}
\caption{Comparison of present model estimation with experimental radius time curves \citep{vanderMeer2007,Doinikov2009,DEJONG2007} for (\textit{a}) $1.7\,\mu$m radius bubble with Gaussian tapered 8-cycle acoustic pulse at $p_a=0.04\,$MPa and $f=2.5\,$MHz (\textit{b}) $1.4\, \mu$m radius bubble steady state response at $p_a=0.1\,$MPa and $f=3\,$MHz (\textit{c}) $1\,\mu$m radius bubble with 6-cycle sine-wave burst at $p_a=0.1\,$MPa and $f=1.8\,$MHz.}
\label{f:3}
\end{figure}

For $1.7\,\mu$m radius bubble we fitted the experimental radius-time data \citep{vanderMeer2007} with the Gaussian tapered 8-cycle acoustic pulse (as assumed in the experiments) of a bubble from the present model at $p_a=0.04\,$MPa and $f=2.5\,$MHz as shown in figure~\ref{f:3}(\textit{a}).
The results shown in figure~\ref{f:3}(\textit{a}) illustrate that the present model provides the best fit with the experimental data in the central region and shows deviations at the beginning and end stages, which is in good agreement with results reported by \citet{Tu2009}. For $1.4\,\mu$m radius bubble, the experimental radius-time data is fitted with the steady-state response of a bubble from the present model using these optimized interface and material parameters at $p_a=0.1\,$MPa and $f=3\,$MHz as shown in figure~\ref{f:3}(\textit{b}). For $1\,\mu$m radius bubble, the experimental radius-time data is fitted with the 6-cycle sine-wave burst at $p_a=0.1\,$MPa and $f=1.8\,$MHz as depicted in figure~\ref{f:3}(\textit{c}).
The optimized values of interface parameters represented in table~\ref{tb:4} correspond to the bubble undergoing expansion from fictitious to natural configuration. In line with the conclusion drawn earlier for a $2\,\mu$m bubble, we observed the possibility of negative dynamic inner interface tension for these cases as well.
\begin{table}
\begin{center}
\begin{tabular}{cccccccccccc}
~ & \multicolumn{10}{c}{{\bf Interface parameters} $\gamma_{ij}\left(\times10^{-2}\right)$}\\
$R_{20}$ \quad & $\gamma_{11}$ \quad & $\gamma_{21}$ \quad & $\gamma_{12}$ \quad & $\gamma_{22}$ \quad & $\gamma_{13}$ \quad & $\gamma_{23}$ \quad & $\gamma_{14}$ \quad & $\gamma_{24}$ \quad & $\gamma_{15}$ \quad & $\gamma_{25}$\\
1.7 \quad & 1.14 \quad & 1.00 \quad & 6.00 \quad & 41.00 \quad & 5.90 \quad & 54.46 \quad & 2.00 \quad & 5.00 \quad & 79.10 \quad & 1.00 \\
1.4 \quad & 1.00 \quad & 1.00 \quad & 1.38 \quad & 49.53 \quad & 1.00 \quad & 65.00 \quad & 1.00 \quad & 1.00 \quad & 80.00 \quad & 1.00 \\
1.0 \quad & 1.54 \quad & 2.20 \quad & 1.00 \quad & 75.36 \quad & 1.00 \quad & 48.32 \quad & 2.00 \quad & 2.00 \quad & 78.00 \quad & 1.00\\
\end{tabular}
\caption{Optimized interface parameters (IP) for a bubble with natural configuration outer radius $R_{20}=1.7\,\mu$m, $1.4\,\mu$m and $1\;\mu$m.}
\label{tb:4}
\end{center}
\end{table}
\begin{table}
\begin{center}
\begin{tabular}{cccccccccc}
~ & \multicolumn{9}{c}{\bf Bulk material parameters}\\
$R_{20}$ \quad & $C_1$ \quad & $C_2$ \quad & $\eta^{\rm L}$ \quad & $\eta^{\rm S}$ \quad & $f$ \quad & $p_{g_0}$ \quad & $h$ \quad & $\zeta$ \quad & $\zeta_0$ \\ 
1.7 \quad & 3.24 \quad & 4.55 \quad & 2.00 \quad & 0.04 \quad & 2.5 \quad & 0.08 \quad & 4.00 \quad & 0.91 \quad & 1.08\\
1.4 \quad & 3.23 \quad & 4.46 \quad & 1.00 \quad & 0.04 \quad & 3.0 \quad & 0.09 \quad & 4.07 \quad & 0.68 \quad & 1.10\\
1.0 \quad & 3.33 \quad & 4.47 \quad & 1.00 \quad & 0.03 \quad & 1.8 \quad & 0.07 \quad & 4.05 \quad & 0.80 \quad & 1.67\\
\end{tabular}
\caption{Optimized material parameters ({MP}) such as shell elastic constants $\left(C_1,C_2\right)\;$MPa, viscosity of shell $\left(\eta^{\rm S}\right)$ Pa-s and liquid $\left(\eta^{\rm L}\right)\;$mPa-s, frequency $\left(f\right)\;$MHz, natural configuration gas pressure $\left(p_{g_0}\right)\;$MPa, bubble thickness $\left(h\right)\;$nm, compression ratios $\zeta$ and $\zeta_0$ for bubble with natural configuration outer radius $R_{20}=1.7\,\mu$m, $1.4\;\mu$m and $1\;\mu$m.}
\label{tb:5}
\end{center}
\end{table}

\citet{DEJONG199295} emphasized that a bubble encapsulated/surrounded/coated with an elastic shell behaves differently from that of a bubble without encapsulation. They also pointed out that the bubble shell is characterized by a shell parameter given by $S_p=Eh/(1-\nu)$, where $E$ is the Young's modulus, $\nu$ is the Poisson's ratio, and $h$ is the shell thickness. Subsequently, the well known Marmottant model \citep{marmottant2005} introduced the shell elasticity modulus, $\chi=S_p/2$, and studied the bubble behavior by taking into consideration the physical properties of the bubble coating. The shell elasticity modulus $(\chi)$ can be related to the shear modulus $(\mu)$ under thin-shelled elasticity theory as $\chi \rm{(N/m)}=3\,\mu\textit{h}$ \citep{boal2002mechanics,Tu2009,Li_2013}. This expression can also be obtained by substituting $E=2\mu(1+\nu)$ and $\nu=0.5$ (assuming to be rubber-like materials) in the expression of $S_p$. In the present model, the shell elastic constants $(C_1,C_2)$ from the Mooney-Rivlin material model are related to the shear modulus $(\mu)$ by the relation $\mu=2(C_1+C_2)$. Similarly the dilatational viscosity of the shell $(k^{\rm S})$ can be expressed in terms of viscosity of the shell $(\eta^{\rm S})$ and the shell thickness $(h)$ as $k^{\rm S}\rm{(kg/s)}=3\,\eta^{\rm S}\textit{h}$ \citep{Tu2009}.

In the available literature, researchers have reported that the viscoelastic properties of the EBs are dependent on their initial size of the EB. Motivated by these observations, we estimate the interface $(\gamma_{ij},i=1,2,j=1\,\rm{to}\;5)$ and material $(C_1,C_2,\eta^{\rm S},p_{g_0},h)$ parameters for various natural configuration outer radii $(R_{20})$ of the bubble ranging from $0.8$ to $3.25\,\mu$m at $p_a=0.15\,$MPa and $f=2.5\,$MHz, tabulated in the supplementary material. In these simulations, we modify the constraints for $\zeta$ and $R_{2}^{\rm min}$ and set it close to their calculated values at the respective bubble radii using the linearized Marmottant model to be in good agreement with \citet{Tu2009} experimental time series data.

\begin{figure}
\centering
\includegraphics[width=0.495\linewidth,trim=0in 0in 1.5in 0in,clip=true]{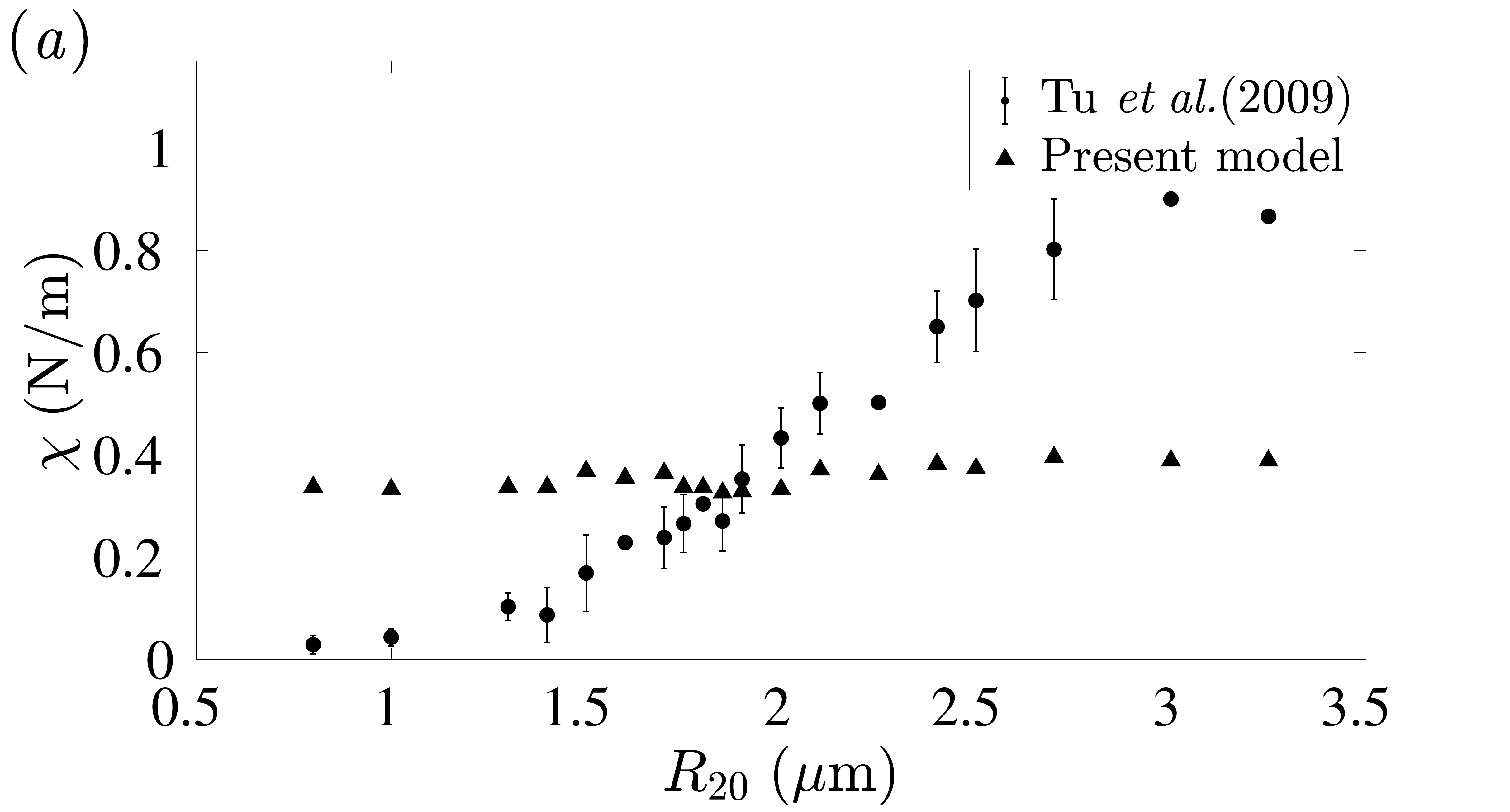}
\includegraphics[width=0.495\linewidth,trim=0in 0in 1.5in 0in,clip=true]{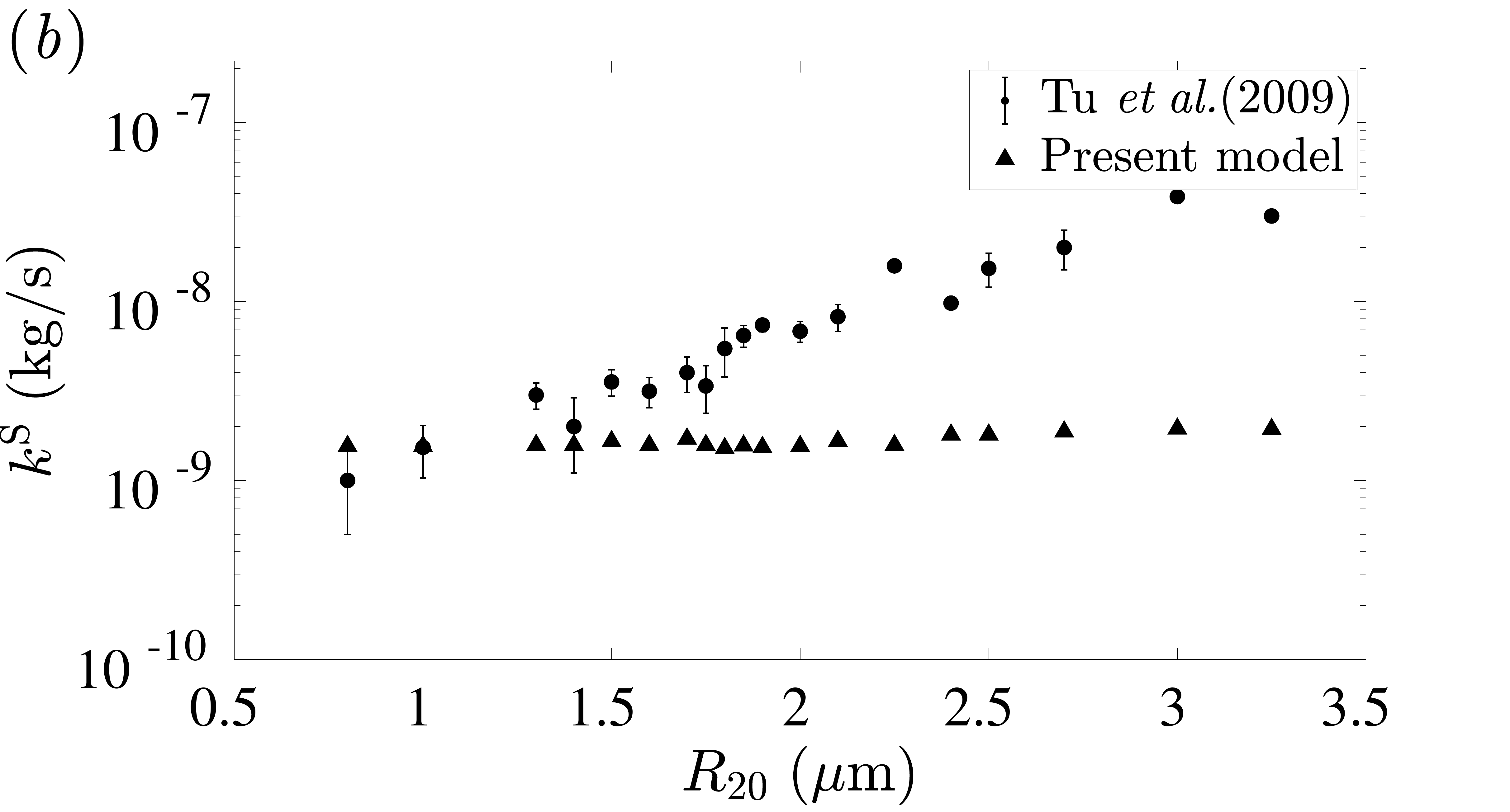}
\caption{Calculated values of shell elasticity parameters $(\chi)$\;N/m and shell dilatational viscosity $(k^{\rm S})\,$kg/s from the optimized interface and material parameters for various natural configuration outer radii $(R_{20})$ of a bubble ranging from $0.8$ to $3.25\,\mu$m at $p_a=0.15\,$MPa and $f=2.5\,$MHz, plotted along with the experimental fitting reported by \citet{Tu2009}.}
\label{f:4}
\end{figure}
\begin{figure}
\centering
\includegraphics[width=0.495\linewidth,trim=0in 0in 1.5in 0in,clip=true]{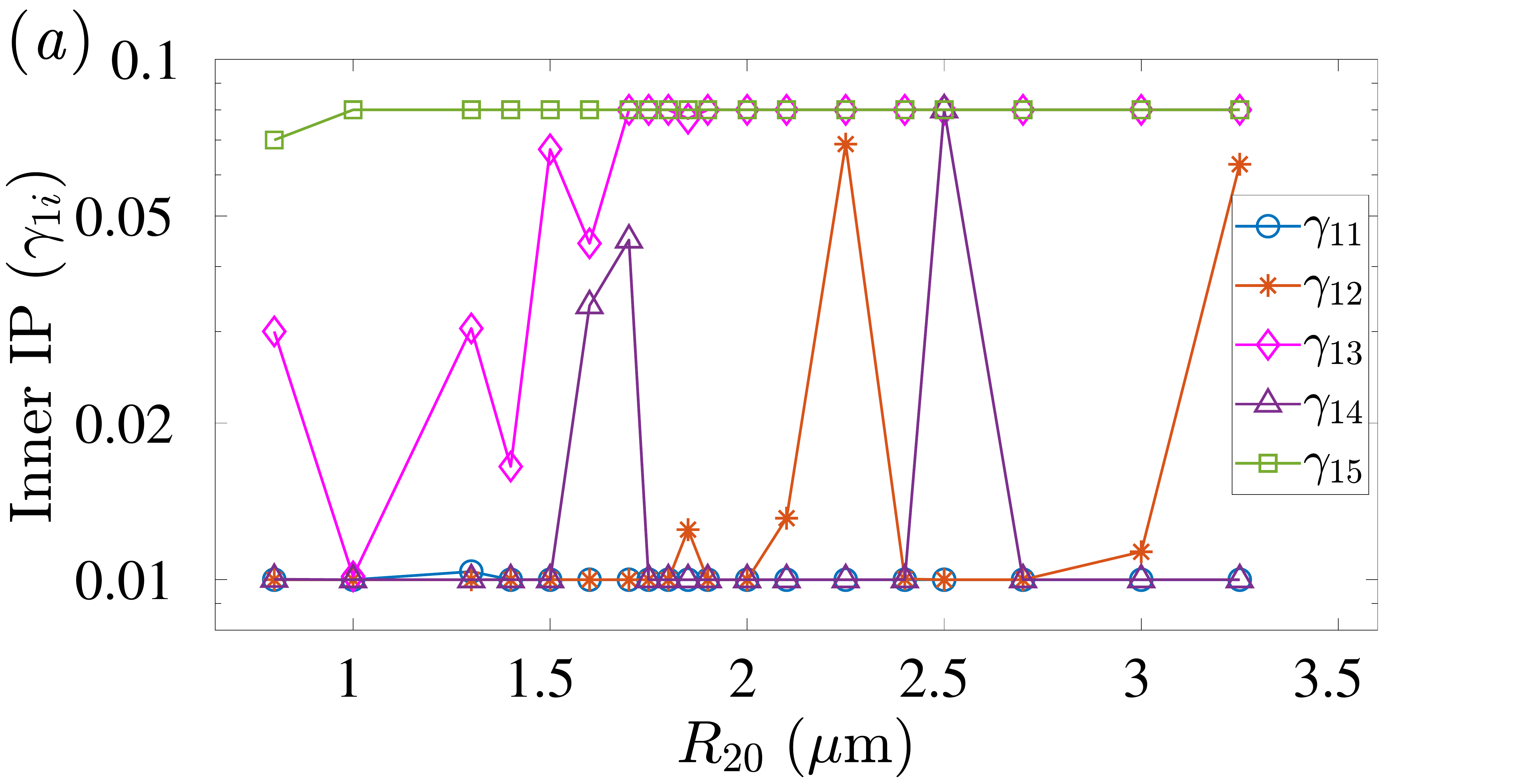}
\includegraphics[width=0.495\linewidth,trim=0in 0in 1.5in 0in,clip=true]{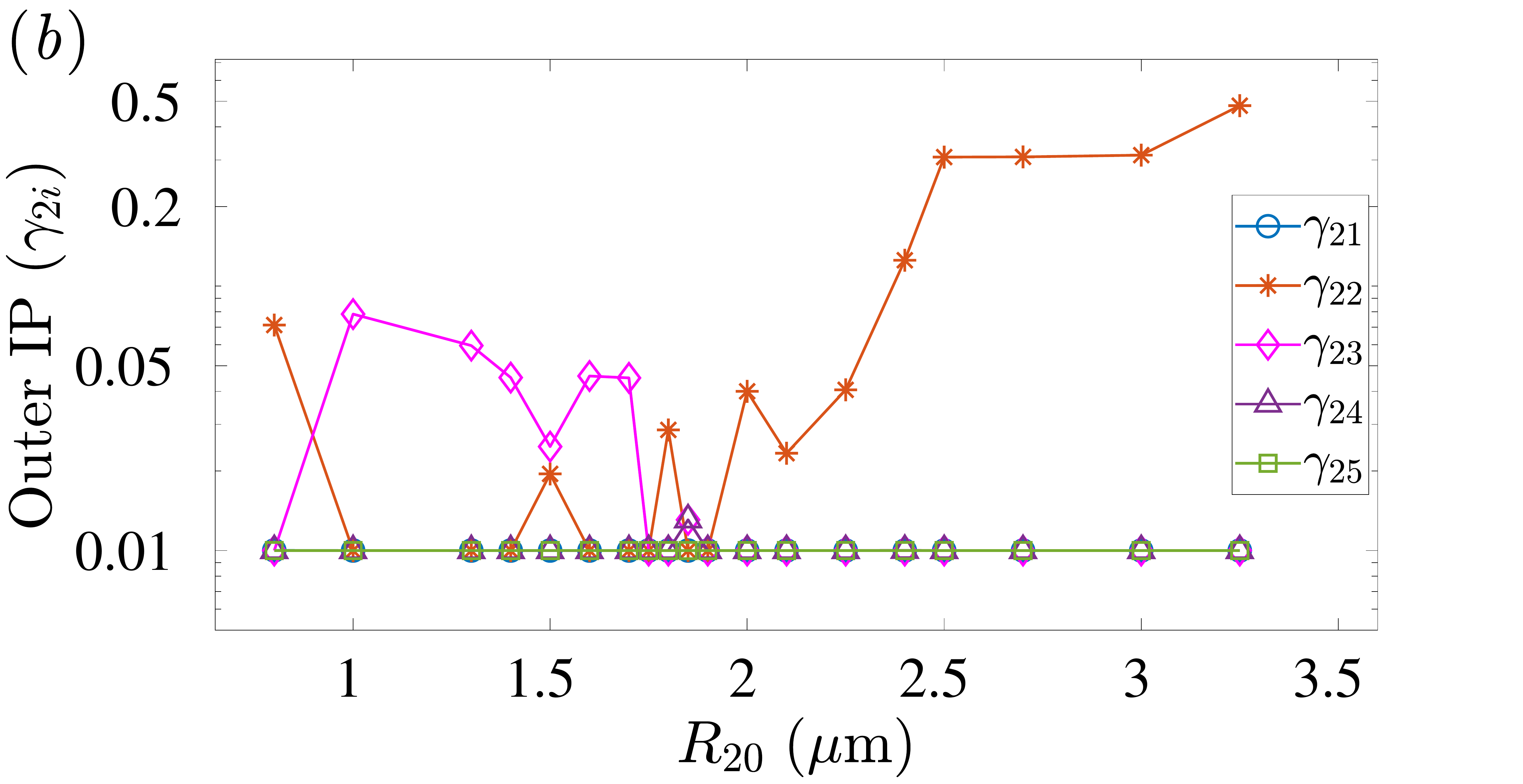}\\
\includegraphics[width=0.495\linewidth,trim=0in 0in 1.5in 0in,clip=true]{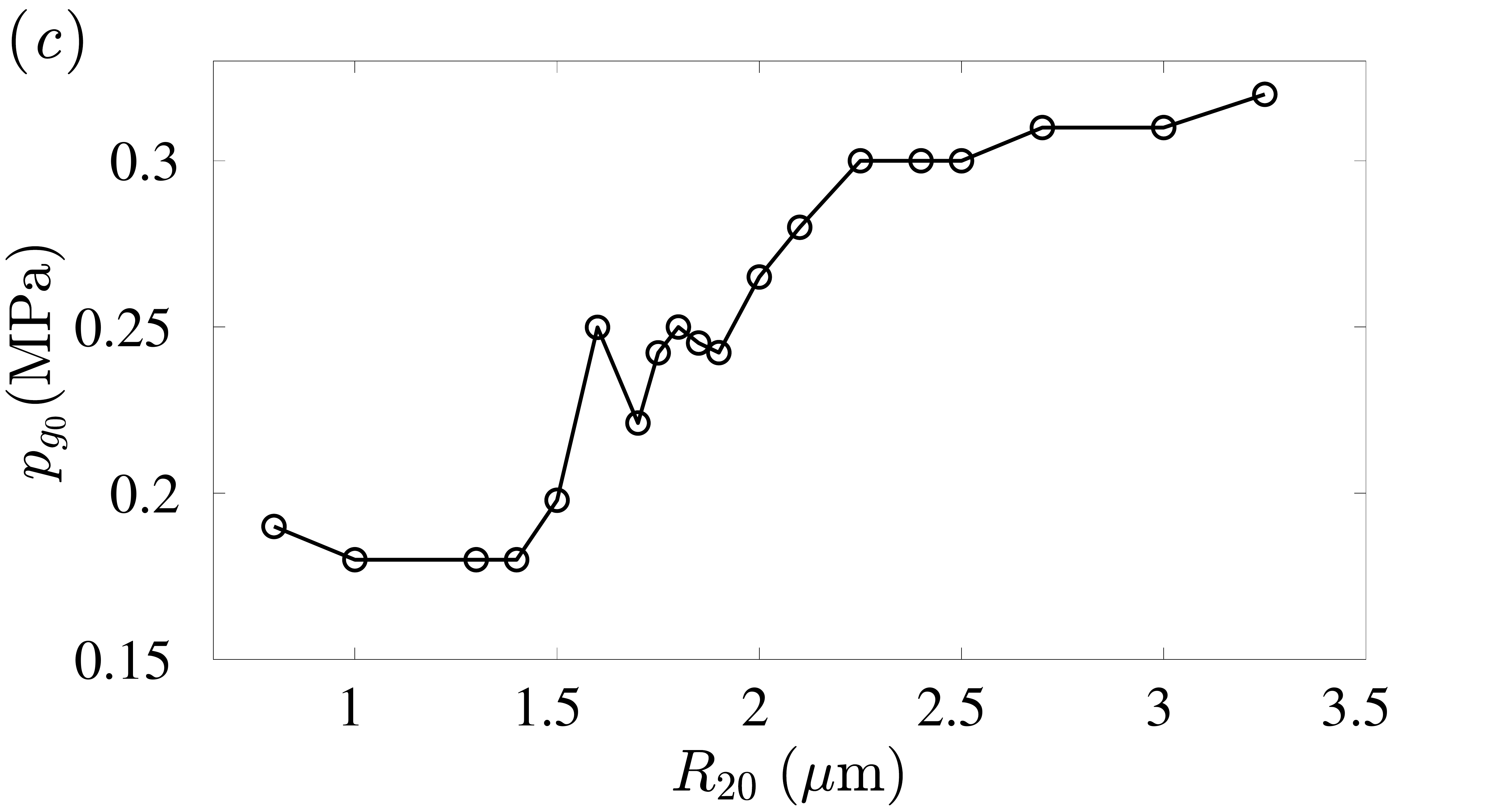}
\caption{Estimated values of (\textit{a}) inner $(\gamma_{1i},i=1\;\rm{to}\;5)$ interface parameters (\textit{b}) outer $(\gamma_{2i},i=1\;\rm{to}\;5)$ interface parameters (\textit{c}) natural configuration gas pressure $(p_{g_0})\,$MPa with the natural configuration outer radii $R_{20}\,(\mu$m) of the bubble from $0.8$ to $3.25\,\mu$m at $p_a=0.15\,$MPa and $f=2.5\,$MHz for the experimental data  \citet{Tu2009}.}
\label{f:5}
\end{figure}

The calculated values of $\chi$ and $k^{\rm S}$ from the optimized interface and material parameters for their respective $R_{20}$ have been plotted along with experimental fitting reported by \citet{Tu2009} in figure~\ref{f:4}. One can see that for different $R_{20}$, the respective values of $\chi$ and $k^{\rm S}$ are almost the same, providing a better fit than that of the linearized Marmottant model. The corresponding estimated values of inner and outer interface parameters and natural configuration pressure are shown in figures~\ref{f:5}(\textit{a}), \ref{f:5}(\textit{b}) and \ref{f:5}(\textit{c}), respectively. As mentioned earlier, the interface parameters here carry the orders discussed at the end of Section \ref{sec:cr}. The bubble shelled models developed earlier, where analysis was carried out without interface energy, the shell parameters were highly dependent on the initial size of the EBs. In the present model, we have the interface parameters which significantly contribute to the mechanics of these EBs through the deformation dependent interface area effects $(\gamma_{i2})$, curvature effects $(\gamma_{i3})$, coupled effects between interface area and curvature $(\gamma_{i5})$, and initial size-dependent effects $(\gamma_{i4})$. These interface parameters allow the present model to eliminate the dependency on the initial radii of the bubble, unlike the observations of \citet{Tu2009}. It is interesting to note that both the shell elastic constants and the interface parameters influence the determination of the other. In figure~\ref{f:5}(\textit{a}), one can see that $\gamma_{15}$ and $\gamma_{13}$ take higher values compared to rest of the interface parameters. As $\gamma_{15}$ contributes only to the prestress at the natural configuration, one can say that interface curvature effects play an important role in the analysis of EBs.

As mentioned earlier, apart from the interface parameters, natural configuration gas pressure $p_{g_0}$ has an important role in estimating the material parameters. For example, using present model, the estimated $p_{g_0}$ value for $1.7\mu$m bubble from two different experiments by \cite{vanderMeer2007} and \cite{Tu2009} are $0.08$ MPa and $0.22$ MPa, respectively, while the estimated viscoelastic constants do not have much difference. This difference in the $p_{g_0}$ values for the same natural configuration radius can be explained through the way natural configuration is attained from the fictitious configuration. It is important to identify that only the estimated shear and viscous shell parameters are close to each other for these two different experiments. The present model also gives a good estimate of $p_{g_0}$ value with radius $R_{20}$ for the experiment by \cite{Tu2009}. The trend obtained in figure~\ref{f:5}(\textit{c}) can be understood from the term $-(\gamma_{15}+\gamma_{13})/R_{10}^2$ in \eqref{e:pg0}. As $\gamma_{15}$ and $\gamma_{13}$ are dominant in figure~\ref{f:5}(\textit{a}), at lower radius, $p_{g_0}$ value is reduced by $-(\gamma_{15}+\gamma_{13})/R_{10}^2$ term as almost all the estimated interface and material parameters for different radii are close. However, with the increase in the radius $R_{20}$, the dominance of these interface parameters reduces, including that of $\gamma_{15}$ and $\gamma_{13}$, and a close to constant trend in $p_{g_0}$ value is observed.

While solving the optimization problem, major care has been taken to avoid getting unrealistically high or low values for any of these optimized parameters. One can further improve the estimation of these interface and material parameters from the generalized framework presented in this paper by modifying the mathematical structure of the Young-Laplace equation \eqref{e:YL} as discussed earlier, along with better viscous models.



\section{Conclusion}
\label{sec:conclusion}
In this work, a simple and generalized interface energy model based on continuum mechanics is introduced to describe the phenomenon associated with encapsulated bubble dynamics. For an encapsulated bubble suspended in a fluid, the Steigmann-Ogden interface between the gas-encapsulation and encapsulation-liquid carries sufficient surface/interface energy that significantly affects the mechanics and cannot be neglected. Notably, the present model considers these interface effects through interface strain and bending rigidity and makes this a robust model over the existing literature. The requirement of fictitious and natural configurations introduced by interface effects in the present model helps understand the physical aspects of encapsulated bubble dynamics. It has been shown that the negative dynamic interface tension associated with interface bending rigidity is favorable for the encapsulated bubble dynamics. Using constrained optimization based on the compression/expansion ratio of the bubble, a better fit of the experimental data is obtained by estimating the interface and material parameters for three different bubble radii. Shell viscoelastic parameters are also estimated for encapsulated bubbles of different radii and obtained a better fit over the existing ones. The present interface energy model is, for the first time, able to resolve the issue of significant dependency of shell viscoelastic properties on the initial radii/size of the encapsulated bubbles.

Important future work will be the study of non-spherical oscillations of such encapsulated bubbles using the present model. It would also be interesting to extend the present model with other viscoelastic models to study the mechanics of encapsulated microbubbles. It would be interesting to explore the influence of different nonlinear interface constitutive relations within the framework presented here. This work can also be generalized to study the problems where interface effects play an essential role. 

\section*{Acknowledgements}
We are grateful to P. G. Senapathy Center for Computing Resource, IIT Madras, for providing the computational resources. This research was supported by the Department of Science and Technology (DST), Government of India, through the INSPIRE faculty research grant under the project number DST/INSPIRE/04/2015/002112. 

\appendix
\section{Calculation of velocity field}
\label{app:A}
Due to the incompressibility of the encapsulation, the shell or material volume is conserved. Hence, we have the following three relations
\begin{align}
    \f{4}{3}\pi(R_2^3(t)-R_1^3(t))&=\f{4}{3}\pi(R_{20}^3-R_{10}^3),\label{appe:comp1}\\
    r^3-R_1^3(t)&=r_e^3-R_{e_1}^3,\label{appe:comp2}\\
    R_2^3(t)-r^3&=R_{e_2}^3-r_e^3\label{appe:comp3}.
\end{align}
Differentiating \eqref{appe:comp1} with respect to $t$, we get
\begin{align}
    R_2^2 \dot R_2&=R_1^2 \dot R_1.
\end{align}
The velocity field at any instant of time as a function of $r$ can be obtained by differentiating \eqref{appe:comp2} with respect to $t$ and expressed as
\begin{align}
    u(r,t)=\f{{\rm d}r}{{\rm d}t}=\f{R_1^2\dot{R}_1}{r^2}=\f{R_2^2\dot{R}_2}{r^2}.
\end{align}

\section{Derivation of the elastic restoring force $(p_e)$}
\label{app:B}

Differentiating the definition of stretch ratio $\Lambda=r/r_e$ and using the relation \eqref{e:incomp} to eliminate $r_e$, we have 
\begin{align}
    \f{{\rm d}r}{r}=\f{{\rm d}\Lambda}{\Lambda(1-\Lambda^3)}.
    \label{appe:dr}
\end{align}
The first term on the right hand side of \eqref{e:inteq} can be simplified using the expression for stresses in the shell as 
\begin{align}
    \int_{R_1}^{R_2} \left(\f{2T_{rr}^{\rm S}-T^{\rm S}_{\phi\phi}-T^{\rm S}_{\theta\theta}}{r}\right) {\rm d}r=&-4\int_{R_1}^{R_2} \bigg[C_1\left(\Lambda^2-\Lambda^{-4}\right)+C_2\left(\Lambda^4-\Lambda^{-2}\right)\bigg]\f{{\rm d}r}{r} \nn \\ 
    &+\int_{R_1}^{R_2}\f{4\eta^{\rm S}}{r}\left(\f{\del u^{\rm S}}{\del r}-\f{u^{\rm S}}{r}\right){\rm d}r.
    \label{appe:st}
\end{align}
Here, the first term on the right hand side of \eqref{appe:st} corresponds to the elastic restoring force $(p_e)$. Using the relation \eqref{appe:dr} and rewriting the integrand of $p_e$ in terms of $\Lambda$, with $\Lambda_1=R_1/R_{e_1}$ and $\Lambda_2=R_2/R_{e_2}$, the expression for the restoring force is calculated as 
\begin{align}
    p_e=&-4\int_{\Lambda_1}^{\Lambda_2} \bigg[C_1\left(\Lambda^2-\Lambda^{-4}\right)+C_2\left(\Lambda^4-\Lambda^{-2}\right)\bigg]\f{{\rm d}\Lambda}{\Lambda(1-\Lambda^3)} \nn \\
    =&-4\left[\f{C_1}{\Lambda}+\f{C_2}{2\Lambda^2}+\f{C_1}{4\Lambda^4}-C_2\Lambda\right]_{\Lambda_2}^{\Lambda_1} \nn \\
    =&-4\left[C_1\left(\f{1}{4\Lambda_2^4}+\f{1}{\Lambda_2}\right)+C_2\left(\f{1}{2\Lambda_2^2}-\Lambda_2\right)\right]+4\left[C_1\left(\f{1}{4\Lambda_1^4}+\f{1}{\Lambda_1}\right)+C_2\left(\f{1}{2\Lambda_1^2}-\Lambda_1\right)\right].
\end{align}

\bibliographystyle{jfm}
\bibliography{references}

\begin{thebibliography}{76}
\expandafter\ifx\csname natexlab\endcsname\relax\def\natexlab#1{#1}\fi
\def\au#1{#1} \def\ed#1{#1} \def\yr#1{#1}\def\at#1{#1}\def\jt#1{\textit{#1}}
  \def\bt#1{#1}\def\bvol#1{\textbf{#1}} \def\vol#1{#1} \def\pg#1{#1}
  \def\publ#1{#1}\def\arxiv#1{#1}\def\org#1{#1}\def\st#1{\textit{#1}}

\bibitem[Abou-Saleh {\em et~al.\/}(2013)Abou-Saleh, Peyman, Critchley, Evans \&
  Thomson]{AbouSaleh2013}
{\sc \au{Abou-Saleh, Radwa~H.}, \au{Peyman, Sally~A.}, \au{Critchley, Kevin},
  \au{Evans, Stephen~D.} \& \au{Thomson, Neil~H.}} \yr{2013}  \at{Nanomechanics
  of lipid encapsulated microbubbles with functional coatings}.  \jt{Langmuir}
  \bvol{29}~(12),  \pg{4096--4103}.

\bibitem[Argudo {\em et~al.\/}(2016)Argudo, Bethel, Marcoline \&
  Grabe]{ARGUDO20161619}
{\sc \au{Argudo, David}, \au{Bethel, Neville~P.}, \au{Marcoline, Frank~V.} \&
  \au{Grabe, Michael}} \yr{2016}  \at{Continuum descriptions of membranes and
  their interaction with proteins: Towards chemically accurate models}.
  \jt{Biochim. Biophys. Acta, Biomembr.}  \bvol{1858}~(7, Part B),
  \pg{1619--1634}.

\bibitem[Bian {\em et~al.\/}(2020)Bian, Litvinov \&
  Koumoutsakos]{bian2020bending}
{\sc \au{Bian, Xin}, \au{Litvinov, Sergey} \& \au{Koumoutsakos, Petros}}
  \yr{2020}  \at{Bending models of lipid bilayer membranes: Spontaneous
  curvature and area-difference elasticity}.  \jt{Comput. Methods Appl. Mech.
  Eng.}  \bvol{359},  \pg{112758}.

\bibitem[Boal(2002)]{boal2002mechanics}
{\sc \au{Boal, David}} \yr{2002} {\em Mechanics of the Cell\/}.
  \publ{Cambridge University Press}.

\bibitem[Bruot \& Caupin(2016)]{PRL2016}
{\sc \au{Bruot, Nicolas} \& \au{Caupin, Fr\'ed\'eric}} \yr{2016}  \at{Curvature
  dependence of the liquid-vapor surface tension beyond the tolman
  approximation}.  \jt{Phys. Rev. Lett.}  \bvol{116},  \pg{056102}.

\bibitem[Chabouh {\em et~al.\/}(2021)Chabouh, Dollet, Quilliet \&
  Coupier]{doi:10.1121/10.0003500}
{\sc \au{Chabouh, Georges}, \au{Dollet, Benjamin}, \au{Quilliet, Catherine} \&
  \au{Coupier, Gwennou}} \yr{2021}  \at{Spherical oscillations of encapsulated
  microbubbles: Effect of shell compressibility and anisotropy}.  \jt{J.
  Acoust. Soc. Am.}  \bvol{149}~(2),  \pg{1240--1257}.

\bibitem[Chomas {\em et~al.\/}(2000)Chomas, Dayton, May, Allen, Klibanov \&
  Ferrara]{chomas2000optical}
{\sc \au{Chomas, James~E}, \au{Dayton, Paul~A}, \au{May, Donovan}, \au{Allen,
  John}, \au{Klibanov, Alexander} \& \au{Ferrara, Katherine}} \yr{2000}
  \at{Optical observation of contrast agent destruction}.  \jt{Appl. Phys.
  Lett.}  \bvol{77}~(7),  \pg{1056--1058}.

\bibitem[Church(1995)]{Church1995}
{\sc \au{Church, Charles~C.}} \yr{1995}  \at{The effects of an elastic solid
  surface layer on the radial pulsations of gas bubbles}.  \jt{J. Acoust. Soc.
  Am.}  \bvol{97}~(3),  \pg{1510--1521}.

\bibitem[Doinikov \& Bouakaz(2011)]{Doinikov2011}
{\sc \au{Doinikov, A~A} \& \au{Bouakaz, A}} \yr{2011}  \at{Review of shell
  models for contrast agent microbubbles}.  \jt{IEEE Trans. Ultrason.
  Ferroelectr. Freq. Control}  \bvol{58}~(5),  \pg{981--993}.

\bibitem[Doinikov {\em et~al.\/}(2009{\natexlab{{\em a\/}}})Doinikov, Haac \&
  Dayton]{Doinikov2009}
{\sc \au{Doinikov, Alexander~A.}, \au{Haac, Jillian~F.} \& \au{Dayton,
  Paul~A.}} \yr{2009{\natexlab{{\em a\/}}}}  \at{Modeling of nonlinear viscous
  stress in encapsulating shells of lipid-coated contrast agent microbubbles}.
  \jt{Ultrasonics}  \bvol{49}~(2),  \pg{269--275}.

\bibitem[Doinikov {\em et~al.\/}(2009{\natexlab{{\em b\/}}})Doinikov, Haac \&
  Dayton]{DOINIKOV2009263}
{\sc \au{Doinikov, Alexander~A.}, \au{Haac, Jillian~F.} \& \au{Dayton,
  Paul~A.}} \yr{2009{\natexlab{{\em b\/}}}}  \at{Resonance frequencies of
  lipid-shelled microbubbles in the regime of nonlinear oscillations}.
  \jt{Ultrasonics}  \bvol{49}~(2),  \pg{263 -- 268}.

\bibitem[Dollet {\em et~al.\/}(2008)Dollet, van~der Meer, Garbin, de~Jong,
  Lohse \& Versluis]{dollet2008nonspherical}
{\sc \au{Dollet, Benjamin}, \au{van~der Meer, Sander~M}, \au{Garbin, Valeria},
  \au{de~Jong, Nico}, \au{Lohse, Detlef} \& \au{Versluis, Michel}} \yr{2008}
  \at{Nonspherical oscillations of ultrasound contrast agent microbubbles}.
  \jt{Ultrasound Med. Biol.}  \bvol{34}~(9),  \pg{1465--1473}.

\bibitem[Errico {\em et~al.\/}(2015)Errico, Pierre, Pezet, Desailly, Lenkei,
  Couture \& Tanter]{errico2015ultrafast}
{\sc \au{Errico, Claudia}, \au{Pierre, Juliette}, \au{Pezet, Sophie},
  \au{Desailly, Yann}, \au{Lenkei, Zsolt}, \au{Couture, Olivier} \& \au{Tanter,
  Mickael}} \yr{2015}  \at{Ultrafast ultrasound localization microscopy for
  deep super-resolution vascular imaging}.  \jt{Nat.}  \bvol{527}~(7579),
  \pg{499--502}.

\bibitem[Frinking \& {de Jong}(1998)]{FRINKING1998523}
{\sc \au{Frinking, Peter~J.A.} \& \au{{de Jong}, Nico}} \yr{1998}  \at{Acoustic
  modeling of shell-encapsulated gas bubbles}.  \jt{Ultrasound Med. Biol.}
  \bvol{24}~(4),  \pg{523--533}.

\bibitem[Fung(1977)]{yuan-chengfung1977}
{\sc \au{Fung, Yuan~Cheng}} \yr{1977} {\em A First Course in Continuum
  Mechanics\/}.  \publ{Prentice Hall}.

\bibitem[Gao {\em et~al.\/}(2014)Gao, Huang, Qu \& Fang]{GAO201459}
{\sc \au{Gao, Xiang}, \au{Huang, Zhuping}, \au{Qu, Jianmin} \& \au{Fang,
  Daining}} \yr{2014}  \at{A curvature-dependent interfacial energy-based
  interface stress theory and its applications to nano-structured materials:
  (i) general theory}.  \jt{J. Mech. Phys. Solids}  \bvol{66},  \pg{59--77}.

\bibitem[Gong {\em et~al.\/}(2014)Gong, Cabodi \& Porter]{Gong2014}
{\sc \au{Gong, Yanjun}, \au{Cabodi, Mario} \& \au{Porter, Tyrone~M.}} \yr{2014}
   \at{Acoustic investigation of pressure-dependent resonance and shell
  elasticity of lipid-coated monodisperse microbubbles}.  \jt{Appl. Phys.
  Lett.}  \bvol{104}~(7),  \pg{074103}.

\bibitem[Gorce {\em et~al.\/}(2000)Gorce, Arditi \& Schneider]{GORCE2000}
{\sc \au{Gorce, J.~N.}, \au{Arditi, M.} \& \au{Schneider, M.}} \yr{2000}
  \at{Influence of bubble size distribution on the echogenicity of ultrasound
  contrast agents}.  \jt{Invest. Radiol.}  \bvol{35}~(11),  \pg{661--671}.

\bibitem[Helfield(2019)]{Helfield2019}
{\sc \au{Helfield, Brandon}} \yr{2019}  \at{A review of phospholipid
  encapsulated ultrasound contrast agent microbubble physics}.  \jt{Ultrasound
  Med. Biol.}  \bvol{45}~(2),  \pg{282--300}.

\bibitem[Helfield \& Goertz(2013)]{Helfield2013}
{\sc \au{Helfield, Brandon~L.} \& \au{Goertz, David~E.}} \yr{2013}
  \at{Nonlinear resonance behavior and linear shell estimates for
  {Definity}{\texttrademark} and {MicroMarker}{\texttrademark} assessed with
  acoustic microbubble spectroscopy}.  \jt{J. Acoust. Soc. Am.}
  \bvol{133}~(2),  \pg{1158--1168}.

\bibitem[Helfrich(1973)]{helfrich1973elastic}
{\sc \au{Helfrich, Wolfgang}} \yr{1973}  \at{Elastic properties of lipid
  bilayers: theory and possible experiments}.  \jt{Z. Naturforsch., C: Biosci.}
   \bvol{28}~(11-12),  \pg{693--703}.

\bibitem[Hernot \& Klibanov(2008)]{HERNOT20081153}
{\sc \au{Hernot, Sophie} \& \au{Klibanov, Alexander~L.}} \yr{2008}
  \at{Microbubbles in ultrasound-triggered drug and gene delivery}.  \jt{Adv.
  Drug Delivery Rev.}  \bvol{60}~(10),  \pg{1153--1166}.

\bibitem[Hoff(2001)]{hoff2001acoustic}
{\sc \au{Hoff, Lars}} \yr{2001} {\em Acoustic characterization of contrast
  agents for medical ultrasound imaging\/}.  \publ{Springer Science \& Business
  Media}.

\bibitem[Hoff {\em et~al.\/}(1996)Hoff, Sontum \& Hoff]{584337}
{\sc \au{Hoff, L.}, \au{Sontum, P.C.} \& \au{Hoff, B.}} \yr{1996} Acoustic
  properties of shell-encapsulated, gas-filled ultrasound contrast agents.
  \bt{In {\em 1996 IEEE Ultrason. Sym. Proc.\/}},  \pg{pp. 1441--1444}.

\bibitem[Itskov(2007)]{itskov2007tensor}
{\sc \au{Itskov, M.}} \yr{2007} {\em Tensor algebra and tensor analysis for
  engineers\/}.  \publ{Springer}.

\bibitem[Jiang \& Lu(2008)]{jiang2008size}
{\sc \au{Jiang, Q} \& \au{Lu, HM}} \yr{2008}  \at{Size dependent interface
  energy and its applications}.  \jt{Surf. Sci. Rep.}  \bvol{63}~(10),
  \pg{427--464}.

\bibitem[de~Jong {\em et~al.\/}(1994)de~Jong, Cornet \& Lancée]{DEJONG1994447}
{\sc \au{de~Jong, N.}, \au{Cornet, R.} \& \au{Lancée, C.T.}} \yr{1994}
  \at{Higher harmonics of vibrating gas-filled microspheres. part one:
  simulations}.  \jt{Ultrasonics}  \bvol{32}~(6),  \pg{447 -- 453}.

\bibitem[de~Jong {\em et~al.\/}(2007)de~Jong, Emmer, Chin, Bouakaz, Mastik,
  Lohse \& Versluis]{DEJONG2007}
{\sc \au{de~Jong, Nico}, \au{Emmer, Marcia}, \au{Chin, Chien~Ting},
  \au{Bouakaz, Ayache}, \au{Mastik, Frits}, \au{Lohse, Detlef} \& \au{Versluis,
  Michel}} \yr{2007}  \at{``compression-only'' behavior of phospholipid-coated
  contrast bubbles}.  \jt{Ultrasound Med. Biol.}  \bvol{33}~(4),  \pg{653 --
  656}.

\bibitem[de~Jong \& Hoff(1993)]{DEJONG1993175}
{\sc \au{de~Jong, N.} \& \au{Hoff, L.}} \yr{1993}  \at{Ultrasound scattering
  properties of albunex microspheres}.  \jt{Ultrasonics}  \bvol{31}~(3),
  \pg{175 -- 181}.

\bibitem[de~Jong {\em et~al.\/}(1992)de~Jong, Hoff, Skotland \&
  Bom]{DEJONG199295}
{\sc \au{de~Jong, N.}, \au{Hoff, L.}, \au{Skotland, T.} \& \au{Bom, N.}}
  \yr{1992}  \at{Absorption and scatter of encapsulated gas filled
  microspheres: Theoretical considerations and some measurements}.
  \jt{Ultrasonics}  \bvol{30}~(2),  \pg{95 -- 103}.

\bibitem[Klibanov(2006)]{Klibanov2006}
{\sc \au{Klibanov, Alexander~L.}} \yr{2006}  \at{Microbubble contrast agents}.
  \jt{Invest. Radiol.}  \bvol{41}~(3),  \pg{354--362}.

\bibitem[Kooiman {\em et~al.\/}(2014)Kooiman, Vos, Versluis \&
  de~Jong]{KOOIMAN201428}
{\sc \au{Kooiman, Klazina}, \au{Vos, Hendrik~J.}, \au{Versluis, Michel} \&
  \au{de~Jong, Nico}} \yr{2014}  \at{Acoustic behavior of microbubbles and
  implications for drug delivery}.  \jt{Adv. Drug Delivery Rev.}  \bvol{72},
  \pg{28--48}.

\bibitem[Li {\em et~al.\/}(2013{\natexlab{{\em a\/}}})Li, Matula, Tu, Guo \&
  Zhang]{Li2013}
{\sc \au{Li, Qian}, \au{Matula, Thomas~J}, \au{Tu, Juan}, \au{Guo, Xiasheng} \&
  \au{Zhang, Dong}} \yr{2013{\natexlab{{\em a\/}}}}  \at{Modeling complicated
  rheological behaviors in encapsulating shells of lipid-coated microbubbles
  accounting for nonlinear changes of both shell viscosity and elasticity}.
  \jt{Phys. Med. Biol.}  \bvol{58}~(4),  \pg{985--998}.

\bibitem[Li {\em et~al.\/}(2013{\natexlab{{\em b\/}}})Li, Matula, Tu, Guo \&
  Zhang]{Li_2013}
{\sc \au{Li, Qian}, \au{Matula, Thomas~J}, \au{Tu, Juan}, \au{Guo, Xiasheng} \&
  \au{Zhang, Dong}} \yr{2013{\natexlab{{\em b\/}}}}  \at{Modeling complicated
  rheological behaviors in encapsulating shells of lipid-coated microbubbles
  accounting for nonlinear changes of both shell viscosity and elasticity}.
  \jt{Phys. Med. Biol.}  \bvol{58}~(4),  \pg{985--998}.

\bibitem[Liang {\em et~al.\/}(2018)Liang, Cao, Wang \& Dobrynin]{ACSletter}
{\sc \au{Liang, Heyi}, \au{Cao, Zhen}, \au{Wang, Zilu} \& \au{Dobrynin,
  Andrey~V.}} \yr{2018}  \at{Surface stress and surface tension in polymeric
  networks}.  \jt{ACS Macro Lett.}  \bvol{7}~(1),  \pg{116--121}.

\bibitem[Lindner(2004)]{Lindner2004}
{\sc \au{Lindner, Jonathan~R.}} \yr{2004}  \at{Microbubbles in medical imaging:
  current applications and future directions}.  \jt{Nat. Rev. Drug Discovery}
  \bvol{3}~(6),  \pg{527--533}.

\bibitem[Liu {\em et~al.\/}(2014)Liu, Zhou, Yang, Shen, Wang, Zhang, Zhi \&
  Wu]{Liu2014}
{\sc \au{Liu, Baoxia}, \au{Zhou, Xiao}, \au{Yang, Fei}, \au{Shen, Hong},
  \au{Wang, Shenguo}, \au{Zhang, Bo}, \au{Zhi, Guang} \& \au{Wu, Decheng}}
  \yr{2014}  \at{Fabrication of uniform sized polylactone microcapsules by
  premix membrane emulsification for ultrasound imaging}.  \jt{Polym. Chem.}
  \bvol{5}~(5),  \pg{1693--1701}.

\bibitem[Liu {\em et~al.\/}(2006)Liu, Miyoshi \& Nakamura]{LIU200689}
{\sc \au{Liu, Yiyao}, \au{Miyoshi, Hirokazu} \& \au{Nakamura, Michihiro}}
  \yr{2006}  \at{Encapsulated ultrasound microbubbles: Therapeutic application
  in drug/gene delivery}.  \jt{J. Controlled Release}  \bvol{114}~(1),
  \pg{89--99}.

\bibitem[Lum {\em et~al.\/}(2016)Lum, Dove, Murray \& Borden]{Lum2016}
{\sc \au{Lum, Jordan~S.}, \au{Dove, Jacob~D.}, \au{Murray, Todd~W.} \&
  \au{Borden, Mark~A.}} \yr{2016}  \at{Single microbubble measurements of lipid
  monolayer viscoelastic properties for small-amplitude oscillations}.
  \jt{Langmuir}  \bvol{32}~(37),  \pg{9410--9417}.

\bibitem[Marmottant {\em et~al.\/}(2005)Marmottant, van~der Meer, Emmer,
  Versluis, de~Jong, Hilgenfeldt \& Lohse]{marmottant2005}
{\sc \au{Marmottant, Philippe}, \au{van~der Meer, Sander}, \au{Emmer, Marcia},
  \au{Versluis, Michel}, \au{de~Jong, Nico}, \au{Hilgenfeldt, Sascha} \&
  \au{Lohse, Detlef}} \yr{2005}  \at{A model for large amplitude oscillations
  of coated bubbles accounting for buckling and rupture}.  \jt{J. Acoust. Soc.
  Am.}  \bvol{118}~(6),  \pg{3499--3505}.

\bibitem[Medasani \& Vasiliev(2009)]{medasani2009computational}
{\sc \au{Medasani, Bharat} \& \au{Vasiliev, Igor}} \yr{2009}  \at{Computational
  study of the surface properties of aluminum nanoparticles}.  \jt{Surf. Sci.}
  \bvol{603}~(13),  \pg{2042--2046}.

\bibitem[van~der Meer {\em et~al.\/}(2007)van~der Meer, Dollet, Voormolen,
  Chin, Bouakaz, de~Jong, Versluis \& Lohse]{vanderMeer2007}
{\sc \au{van~der Meer, Sander~M.}, \au{Dollet, Benjamin}, \au{Voormolen,
  Marco~M.}, \au{Chin, Chien~T.}, \au{Bouakaz, Ayache}, \au{de~Jong, Nico},
  \au{Versluis, Michel} \& \au{Lohse, Detlef}} \yr{2007}  \at{Microbubble
  spectroscopy of ultrasound contrast agents}.  \jt{J. Acoust. Soc. Am.}
  \bvol{121}~(1),  \pg{648--656}.

\bibitem[Moody \& Attard(2003)]{PRL2003}
{\sc \au{Moody, Michael~P.} \& \au{Attard, Phil}} \yr{2003}
  \at{Curvature-dependent surface tension of a growing droplet}.  \jt{Phys.
  Rev. Lett.}  \bvol{91},  \pg{056104}.

\bibitem[Morgan {\em et~al.\/}(2000)Morgan, Allen, Dayton, Chomas, Klibaov \&
  Ferrara]{Morgan2000}
{\sc \au{Morgan, K.E.}, \au{Allen, J.S.}, \au{Dayton, P.A.}, \au{Chomas, J.E.},
  \au{Klibaov, A.L.} \& \au{Ferrara, K.W.}} \yr{2000}  \at{Experimental and
  theoretical evaluation of microbubble behavior: effect of transmitted phase
  and bubble size}.  \jt{IEEE Trans. Ultrason. Ferroelectr. Freq.}
  \bvol{47}~(6),  \pg{1494--1509}.

\bibitem[Parrales {\em et~al.\/}(2014)Parrales, Fernandez, Perez-Saborid,
  Kopechek \& Porter]{Parrales2014}
{\sc \au{Parrales, Miguel~A.}, \au{Fernandez, Juan~M.}, \au{Perez-Saborid,
  Miguel}, \au{Kopechek, Jonathan~A.} \& \au{Porter, Tyrone~M.}} \yr{2014}
  \at{Acoustic characterization of monodisperse lipid-coated microbubbles:
  Relationship between size and shell viscoelastic properties}.  \jt{J. Acoust.
  Soc. Am.}  \bvol{136}~(3),  \pg{1077--1084}.

\bibitem[Paul {\em et~al.\/}(2010)Paul, Katiyar, Sarkar, Chatterjee, Shi \&
  Forsberg]{Paul2010}
{\sc \au{Paul, Shirshendu}, \au{Katiyar, Amit}, \au{Sarkar, Kausik},
  \au{Chatterjee, Dhiman}, \au{Shi, William~T.} \& \au{Forsberg, Flemming}}
  \yr{2010}  \at{Material characterization of the encapsulation of an
  ultrasound contrast microbubble and its subharmonic response:
  Strain-softening interfacial elasticity model}.  \jt{J. Acoust. Soc. Am.}
  \bvol{127}~(6),  \pg{3846--3857}.

\bibitem[Postema {\em et~al.\/}(2004)Postema, Van~Wamel, Lanc{\'e}e \&
  De~Jong]{postema2004ultrasound}
{\sc \au{Postema, Michiel}, \au{Van~Wamel, Annemieke}, \au{Lanc{\'e}e,
  Charles~T} \& \au{De~Jong, Nico}} \yr{2004}  \at{Ultrasound-induced
  encapsulated microbubble phenomena}.  \jt{Ultrasound Med. Biol.}
  \bvol{30}~(6),  \pg{827--840}.

\bibitem[Qin \& Ferrara(2010)]{Qin2010}
{\sc \au{Qin, Shengping} \& \au{Ferrara, Katherine~W.}} \yr{2010}  \at{A model
  for the dynamics of ultrasound contrast agents in vivo}.  \jt{J. Acoust. Soc.
  Am.}  \bvol{128}~(3),  \pg{1511}.

\bibitem[Rallabandi {\em et~al.\/}(2019)Rallabandi, Marthelot, Jambon-Puillet,
  Brun \& Eggers]{PRL1}
{\sc \au{Rallabandi, Bhargav}, \au{Marthelot, Joel}, \au{Jambon-Puillet,
  Etienne}, \au{Brun, P.~T.} \& \au{Eggers, Jens}} \yr{2019}  \at{Curvature
  regularization near contacts with stretched elastic tubes}.  \jt{Phys. Rev.
  Lett.}  \bvol{123},  \pg{168002}.

\bibitem[Rangamani {\em et~al.\/}(2013)Rangamani, Benjamini, Agrawal, Smit,
  Steigmann \& Oster]{Rangamani2013}
{\sc \au{Rangamani, Padmini}, \au{Benjamini, Ayelet}, \au{Agrawal, Ashutosh},
  \au{Smit, Berend}, \au{Steigmann, David~J.} \& \au{Oster, George}} \yr{2013}
  \at{Small scale membrane mechanics}.  \jt{Biomech. Model. Mechanobiol.}
  \bvol{13}~(4),  \pg{697--711}.

\bibitem[van Rooij {\em et~al.\/}(2015)van Rooij, Luan, Renaud, van~der Steen,
  Versluis, de~Jong \& Kooiman]{vanRooij2015}
{\sc \au{van Rooij, Tom}, \au{Luan, Ying}, \au{Renaud, Guillaume}, \au{van~der
  Steen, Antonius~F.W.}, \au{Versluis, Michel}, \au{de~Jong, Nico} \&
  \au{Kooiman, Klazina}} \yr{2015}  \at{Non-linear response and viscoelastic
  properties of lipid-coated microbubbles: {DSPC} versus {DPPC}}.
  \jt{Ultrasound Med. Biol.}  \bvol{41}~(5),  \pg{1432--1445}.

\bibitem[Sagis(2014)]{Sagis2014}
{\sc \au{Sagis, L. M.~C}} \yr{2014}  \at{Dynamic behavior of interfaces:
  Modeling with nonequilibrium thermodynamics}.  \jt{Adv. Colloid Interface
  Sci.}  \bvol{206},  \pg{328--343}.

\bibitem[Santos {\em et~al.\/}(2012)Santos, Morris, Glynos, Sboros \&
  Koutsos]{BuchnerSantos2012}
{\sc \au{Santos, Evelyn~Buchner}, \au{Morris, Julia~K.}, \au{Glynos,
  Emmanouil}, \au{Sboros, Vassilis} \& \au{Koutsos, Vasileios}} \yr{2012}
  \at{Nanomechanical properties of phospholipid microbubbles}.  \jt{Langmuir}
  \bvol{28}~(13),  \pg{5753--5760}.

\bibitem[Sarkar {\em et~al.\/}(2005)Sarkar, Shi, Chatterjee \&
  Forsberg]{sarkar2005}
{\sc \au{Sarkar, Kausik}, \au{Shi, William~T.}, \au{Chatterjee, Dhiman} \&
  \au{Forsberg, Flemming}} \yr{2005}  \at{Characterization of ultrasound
  contrast microbubbles using in vitro experiments and viscous and viscoelastic
  interface models for encapsulation}.  \jt{J. Acoust. Soc. Am.}
  \bvol{118}~(1),  \pg{539--550}.

\bibitem[Schulman {\em et~al.\/}(2018)Schulman, Trejo, Salez, Rapha{\"e}l \&
  Dalnoki-Veress]{Nat1}
{\sc \au{Schulman, Rafael~D.}, \au{Trejo, Miguel}, \au{Salez, Thomas},
  \au{Rapha{\"e}l, Elie} \& \au{Dalnoki-Veress, Kari}} \yr{2018}  \at{Surface
  energy of strained amorphous solids}.  \jt{Nat. Commun.}  \bvol{9}~(1),
  \pg{982}.

\bibitem[Segers {\em et~al.\/}(2020)Segers, Gaud, Casqueiro, Lassus, Versluis
  \& Frinking]{Segers2020}
{\sc \au{Segers, Tim}, \au{Gaud, Emmanuel}, \au{Casqueiro, Gilles}, \au{Lassus,
  Anne}, \au{Versluis, Michel} \& \au{Frinking, Peter}} \yr{2020}
  \at{Foam-free monodisperse lipid-coated ultrasound contrast agent synthesis
  by flow-focusing through multi-gas-component microbubble stabilization}.
  \jt{Appl. Phys. Lett.}  \bvol{116}~(17),  \pg{173701}.

\bibitem[Segers {\em et~al.\/}(2016)Segers, de~Rond, de~Jong, Borden \&
  Versluis]{Segers2016}
{\sc \au{Segers, Tim}, \au{de~Rond, Leonie}, \au{de~Jong, Nico}, \au{Borden,
  Mark} \& \au{Versluis, Michel}} \yr{2016}  \at{Stability of monodisperse
  phospholipid-coated microbubbles formed by flow-focusing at high production
  rates}.  \jt{Langmuir}  \bvol{32}~(16),  \pg{3937--3944}.

\bibitem[Shafi {\em et~al.\/}(2019)Shafi, McClements, Albaijan, Abou-Saleh,
  Moran \& Koutsos]{Shafi2019}
{\sc \au{Shafi, Adeel~S.}, \au{McClements, Jake}, \au{Albaijan, Ibrahim},
  \au{Abou-Saleh, Radwa~H.}, \au{Moran, Carmel} \& \au{Koutsos, Vasileios}}
  \yr{2019}  \at{Probing phospholipid microbubbles by atomic force microscopy
  to quantify bubble mechanics and nanostructural shell properties}.
  \jt{Colloids Surf., B}  \bvol{181},  \pg{506--515}.

\bibitem[Song {\em et~al.\/}(2018)Song, Peng, Xu, Wang, Yu, Hou, Zou \&
  Yao]{Song2018}
{\sc \au{Song, Ruyuan}, \au{Peng, Chuan}, \au{Xu, Xiaonan}, \au{Wang, Jianwei},
  \au{Yu, Miao}, \au{Hou, Youmin}, \au{Zou, Ruhai} \& \au{Yao, Shuhuai}}
  \yr{2018}  \at{Controllable formation of monodisperse polymer microbubbles as
  ultrasound contrast agents}.  \jt{{ACS} Appl. Mater. Interfaces}
  \bvol{10}~(17),  \pg{14312--14320}.

\bibitem[Steigmann(2001)]{steigmann_2001}
{\sc \au{Steigmann, D.~J.}} \yr{2001} {\em Nonlinear Elasticity: Theory and
  Applications\/}.  \publ{Cambridge University Press}.

\bibitem[Steigmann \& Ogden(1999)]{steigmann}
{\sc \au{Steigmann, D.~J.} \& \au{Ogden, R.~W.}} \yr{1999}  \at{Elastic
  surface-substrate interactions}.  \jt{Proc. R. Soc. London, Ser. A}
  \bvol{455}~(1982),  \pg{437--474}.

\bibitem[Stride \& Saffari(2003)]{Stride2003}
{\sc \au{Stride, E} \& \au{Saffari, N}} \yr{2003}  \at{Microbubble ultrasound
  contrast agents: A review}.  \jt{Proc. Inst. Mech. Eng., Part H: J. Eng.
  Med.}  \bvol{217}~(6),  \pg{429--447}.

\bibitem[Tachibana \& Tachibana(1999)]{1347-4065-38-5S-3014}
{\sc \au{Tachibana, Katsuro} \& \au{Tachibana, Shunro}} \yr{1999}
  \at{Application of ultrasound energy as a new drug delivery system}.
  \jt{Jpn. J. Appl. Phys.}  \bvol{38}~(5S),  \pg{3014}.

\bibitem[Tolman(1949)]{tolman1949effect}
{\sc \au{Tolman, Richard~C}} \yr{1949}  \at{The effect of droplet size on
  surface tension}.  \jt{J. Chem. Phys.}  \bvol{17}~(3),  \pg{333--337}.

\bibitem[Tsiglifis \& Pelekasis(2008)]{tsiglifis2008nonlinear}
{\sc \au{Tsiglifis, Kostas} \& \au{Pelekasis, Nikos~A}} \yr{2008}
  \at{Nonlinear radial oscillations of encapsulated microbubbles subject to
  ultrasound: The effect of membrane constitutive law}.  \jt{J. Acoust. Soc.
  Am.}  \bvol{123}~(6),  \pg{4059--4070}.

\bibitem[Tsutsui {\em et~al.\/}(2004)Tsutsui, Xie \& Porter]{Tsutsui2004}
{\sc \au{Tsutsui, Jeane~M}, \au{Xie, Feng} \& \au{Porter, Richard~Thomas}}
  \yr{2004}  \at{The use of microbubbles to target drug delivery}.
  \jt{Cardiovasc. Ultrasound}  \bvol{2}~(1),  \pg{1--7}.

\bibitem[Tu {\em et~al.\/}(2009)Tu, Guan, Qiu \& Matula]{Tu2009}
{\sc \au{Tu, Juan}, \au{Guan, Jingfeng}, \au{Qiu, Yuanyuan} \& \au{Matula,
  Thomas~J.}} \yr{2009}  \at{Estimating the shell parameters of
  {SonoVue}\textsuperscript{\textregistered} microbubbles using light
  scattering}.  \jt{J. Acoust. Soc. Am.}  \bvol{126}~(6),  \pg{2954--2962}.

\bibitem[Unger {\em et~al.\/}(2001)Unger, Hersh, Vannan, Matsunaga \&
  McCreery]{UNGER200145}
{\sc \au{Unger, Evan~C}, \au{Hersh, Evan}, \au{Vannan, Mani}, \au{Matsunaga,
  Terry~O} \& \au{McCreery, Thomas}} \yr{2001}  \at{Local drug and gene
  delivery through microbubbles}.  \jt{Prog. Cardiovasc. Dis.}  \bvol{44}~(1),
  \pg{45--54}.

\bibitem[Unger {\em et~al.\/}(2004)Unger, Porter, Culp, Labell, Matsunaga \&
  Zutshi]{UNGER20041291}
{\sc \au{Unger, Evan~C.}, \au{Porter, Thomas}, \au{Culp, William}, \au{Labell,
  Rachel}, \au{Matsunaga, Terry} \& \au{Zutshi, Reena}} \yr{2004}
  \at{Therapeutic applications of lipid-coated microbubbles}.  \jt{Adv. Drug
  Delivery Rev.}  \bvol{56}~(9),  \pg{1291--1314}.

\bibitem[{van der Meer} {\em et~al.\/}(2006){van der Meer}, {Dollet}, {Goertz},
  {de Jong}, {Versluis} \& {Lohse}]{4151897}
{\sc \au{{van der Meer}, S.~M.}, \au{{Dollet}, B.}, \au{{Goertz}, D.~E.},
  \au{{de Jong}, N.}, \au{{Versluis}, M.} \& \au{{Lohse}, D.}} \yr{2006}
  Surface modes of ultrasound contrast agent microbubbles.  \bt{In {\em 2006
  IEEE Ultrasonics Symposium\/}},  \pg{pp. 112--115}.

\bibitem[Versluis {\em et~al.\/}(2010)Versluis, Goertz, Palanchon, Heitman,
  van~der Meer, Dollet, de~Jong \& Lohse]{PhysRevE.82.026321}
{\sc \au{Versluis, Michel}, \au{Goertz, David~E.}, \au{Palanchon, Peggy},
  \au{Heitman, Ivo~L.}, \au{van~der Meer, Sander~M.}, \au{Dollet, Benjamin},
  \au{de~Jong, Nico} \& \au{Lohse, Detlef}} \yr{2010}  \at{Microbubble shape
  oscillations excited through ultrasonic parametric driving}.  \jt{Phys. Rev.
  E}  \bvol{82},  \pg{026321}.

\bibitem[Versluis {\em et~al.\/}(2020)Versluis, Stride, Lajoinie, Dollet \&
  Segers]{VERSLUIS20202117}
{\sc \au{Versluis, Michel}, \au{Stride, Eleanor}, \au{Lajoinie, Guillaume},
  \au{Dollet, Benjamin} \& \au{Segers, Tim}} \yr{2020}  \at{Ultrasound contrast
  agent modeling: A review}.  \jt{Ultrasound Med. Biol.}  \bvol{46}~(9),
  \pg{2117--2144}.

\bibitem[{Versluis} {\em et~al.\/}(2004){Versluis}, {van der Meer}, {Lohse},
  {Palanchon}, {Goertz}, {Chin} \& {de Jong}]{1417703}
{\sc \au{{Versluis}, M.}, \au{{van der Meer}, S.~M.}, \au{{Lohse}, D.},
  \au{{Palanchon}, P.}, \au{{Goertz}, D.}, \au{{Chin}, C.~T.} \& \au{{de Jong},
  N.}} \yr{2004} Microbubble surface modes.  \bt{In {\em IEEE Ultrasonics
  Symposium, 2004\/}},  \pg{pp. 207--209}.

\bibitem[Vos {\em et~al.\/}(2011)Vos, Dollet, Versluis \&
  de~Jong]{vos2011nonspherical}
{\sc \au{Vos, Hendrik~J}, \au{Dollet, Benjamin}, \au{Versluis, Michel} \&
  \au{de~Jong, Nico}} \yr{2011}  \at{Nonspherical shape oscillations of coated
  microbubbles in contact with a wall}.  \jt{Ultrasound Med. Biol.}
  \bvol{37}~(6),  \pg{935--948}.

\bibitem[Wang {\em et~al.\/}(2020)Wang, Xue, zhao, Qin, Zhang \& Li]{Wang2020}
{\sc \au{Wang, Qiaozhi}, \au{Xue, Chunlong}, \au{zhao, Hui}, \au{Qin, Yan},
  \au{Zhang, Xiaohan} \& \au{Li, Ying}} \yr{2020}  \at{The fabrication of
  protein microbubbles with diverse gas core and the novel exploration on the
  role of interface introduction in protein crystallization}.  \jt{Colloids
  Surf., A}  \bvol{589},  \pg{124471}.

\bibitem[Zheng(1993)]{zheng1993two}
{\sc \au{Zheng, Q~S}} \yr{1993}  \at{Two-dimensional tensor function
  representation for all kinds of material symmetry}.  \jt{Proc. R. Soc.
  London, Ser. A}  \bvol{443}~(1917),  \pg{127--138}.

\end{thebibliography}

\end{document}


\maketitle

The optimized values of interface and material parameters obtained from the optimization problem for various natural configuration outer radii of the encapsulated bubble are tabulated in tables~\ref{stb:1} and \ref{stb:2}, respectively. The tabulated data below has been used to obtain figures$~4$ and $5$ in the main text.

\begin{table}
\begin{center}
\begin{tabular}{ccccccccccc}
~ & \multicolumn{10}{c}{{\bf Interface parameters} $\gamma_{ij}\left(\times10^{-2}\right)$}\\
$R_{20}$ \quad & $\gamma_{11}$ \quad & $\gamma_{21}$ \quad & $\gamma_{12}$ \quad & $\gamma_{22}$ \quad & $\gamma_{13}$ \quad & $\gamma_{23}$ \quad &  $\gamma_{14}$ \quad & $\gamma_{24}$ \quad & $\gamma_{15}$ \quad & $\gamma_{25}$\\
0.80 \quad & 1.00 \quad & 1.00 \quad & ~1.00 \quad & 12.00 \quad & ~3.00 \quad & 1.00 \quad & 1.00 \quad & 1.00 \quad & 7.00 \quad & 1.00 \\
1.00 \quad & 1.00 \quad & 1.00 \quad & ~1.00 \quad & ~1.00 \quad & ~1.01 \quad & 7.83 \quad & 1.00 \quad & 1.00 \quad & 8.00 \quad & 1.00 \\
1.30 \quad & 1.03 \quad & 1.00 \quad & ~1.00 \quad & ~1.00 \quad & ~3.04 \quad & 5.95 \quad & 1.00 \quad & 1.00 \quad & 8.00 \quad & 1.00 \\
1.40 \quad & 1.00 \quad & 1.00 \quad & ~1.00 \quad & ~1.00 \quad & ~1.65 \quad & 4.51 \quad & 1.00 \quad & 1.00 \quad & 8.00 \quad & 1.00 \\
1.50 \quad & 1.00 \quad & 1.00 \quad & ~1.00 \quad & ~1.94 \quad & ~6.71 \quad & 2.47 \quad & 1.00 \quad & 1.00 \quad & 8.00 \quad & 1.00 \\
1.60 \quad & 1.00 \quad & 1.00 \quad & ~1.00 \quad & ~1.00 \quad & ~4.43 \quad & 4.57 \quad & 3.36 \quad & 1.00 \quad & 8.00 \quad & 1.00 \\
1.70 \quad & 1.10 \quad & 1.00 \quad & ~1.00 \quad & ~1.00 \quad & ~8.00 \quad & 4.49 \quad & 4.49 \quad & 1.00 \quad & 8.00 \quad & 1.00 \\
1.75 \quad & 1.00 \quad & 1.00 \quad & ~1.00 \quad & ~1.00 \quad & ~8.00 \quad & 1.00 \quad & 1.00 \quad & 1.00 \quad & 8.00 \quad & 1.00 \\
1.80 \quad & 1.00 \quad & 1.00 \quad & ~1.00 \quad & ~2.85 \quad & ~8.00 \quad & 1.00 \quad & 1.00 \quad & 1.00 \quad & 8.00 \quad & 1.00 \\
1.85 \quad & 1.00 \quad & 1.00 \quad & ~1.24 \quad & ~1.00 \quad & ~7.69 \quad & 1.30 \quad & 1.00 \quad & 1.30 \quad & 8.00 \quad & 1.00 \\
1.90 \quad & 1.00 \quad & 1.00 \quad & ~1.00 \quad & ~1.00 \quad & ~8.00 \quad & 1.00 \quad & 1.00 \quad & 1.00 \quad & 8.00 \quad & 1.00 \\
2.00 \quad & 1.00 \quad & 1.00 \quad & ~1.00 \quad & ~4.00 \quad & ~8.00 \quad & 1.00 \quad & 1.00 \quad & 1.00 \quad & 8.00 \quad & 1.00 \\
2.10 \quad & 1.00 \quad & 1.00 \quad & ~1.31 \quad & ~2.33 \quad & ~8.00 \quad & 1.00 \quad & 1.00 \quad & 1.00 \quad & 8.00 \quad & 1.00 \\
2.25 \quad & 1.00 \quad & 1.00 \quad & ~6.87 \quad & ~4.05 \quad & ~8.00 \quad & 1.00 \quad & 1.00 \quad & 1.00 \quad & 8.00 \quad & 1.00 \\
2.40 \quad & 1.00 \quad & 1.00 \quad & ~1.00 \quad & 12.52 \quad & ~8.00 \quad & 1.00 \quad & 1.00 \quad & 1.00 \quad & 8.00 \quad & 1.00 \\
2.50 \quad & 1.00 \quad & 1.00 \quad & ~1.00 \quad & 30.75 \quad & ~8.00 \quad & 1.00 \quad & 8.00 \quad & 1.00 \quad & 8.00 \quad & 1.00 \\
2.70 \quad & 1.00 \quad & 1.00 \quad & ~1.00 \quad & 30.81 \quad & ~8.00 \quad & 1.00 \quad & 1.00 \quad & 1.00 \quad & 8.00 \quad & 1.00 \\
3.00 \quad & 1.00 \quad & 1.00 \quad & ~1.13 \quad & 31.28 \quad & ~8.00 \quad & 1.00 \quad & 1.00 \quad & 1.00 \quad & 8.00 \quad & 1.00 \\
3.25 \quad & 1.00 \quad & 1.00 \quad & ~6.28 \quad & 48.12 \quad & ~8.00 \quad & 1.00 \quad & 1.00 \quad & 1.00 \quad & 8.00 \quad & 1.00 \\
\end{tabular}
\caption{Optimized interface parameters ({IP}) for encapsulated bubbles with different natural configuration outer radii $R_{20}\,(\mu$m), excitation pressure $p_a=0.15\,$MPa and frequency $f=2.5\,$MHz.}
\label{stb:1}
\end{center}
\end{table}

\begin{table}
\begin{center}
\begin{tabular}{cccccccc}
~ & \multicolumn{7}{c}{\bf Bulk material parameters}\\
$R_{20}$ \quad & $C_1$ \quad & $C_2$ \quad & $h$ \quad & $\eta^{\rm S}$ \quad & $p_{g_0}$ \quad & $\chi$ \quad & $k^{\rm S}$ \\
0.80 \quad & 2.60 \quad & 4.99 \quad &  7.40 \quad & 0.07 \quad & 0.19 \quad & 0.33 \quad & 1.55 \\
1.00 \quad & 4.43 \quad & 3.06 \quad &  7.40 \quad & 0.07 \quad & 0.18 \quad & 0.33 \quad & 1.55 \\
1.30 \quad & 3.00 \quad & 4.50 \quad &  7.50 \quad & 0.07 \quad & 0.18 \quad & 0.33 \quad &  1.57 \\
1.40 \quad & 4.49 \quad & 3.00 \quad &  7.50 \quad & 0.07 \quad & 0.18 \quad & 0.33 \quad & 1.57 \\
1.50 \quad & 3.46 \quad & 4.32 \quad &  7.87 \quad & 0.07 \quad & 0.19 \quad & 0.36 \quad & 1.65 \\
1.60 \quad & 3.00 \quad & 4.89 \quad &  7.50 \quad & 0.07 \quad & 0.24 \quad & 0.35 \quad & 1.57 \\
1.70 \quad & 4.73 \quad & 3.26 \quad &  7.59 \quad & 0.08 \quad & 0.22 \quad & 0.36 \quad & 1.70 \\
1.75 \quad & 3.03 \quad & 4.46 \quad &  7.50 \quad & 0.07 \quad & 0.24 \quad & 0.33 \quad & 1.57 \\
1.80 \quad & 3.00 \quad & 4.78 \quad &  7.20 \quad & 0.07 \quad & 0.25 \quad & 0.33 \quad & 1.51 \\
1.85 \quad & 3.08 \quad & 4.45 \quad &  7.20 \quad & 0.07 \quad & 0.24 \quad & 0.32 \quad & 1.56 \\
1.90 \quad & 3.02 \quad & 4.47 \quad &  7.30 \quad & 0.07 \quad & 0.24 \quad & 0.32 \quad & 1.53 \\
2.00 \quad & 3.00 \quad & 4.50 \quad &  7.40 \quad & 0.07 \quad & 0.26 \quad & 0.33 \quad & 1.55 \\
2.10 \quad & 3.00 \quad & 4.82 \quad &  7.90 \quad & 0.07 \quad & 0.28 \quad & 0.37 \quad & 1.66 \\
2.25 \quad & 3.03 \quad & 5.00 \quad &  7.49 \quad & 0.07 \quad & 0.30 \quad & 0.36 \quad & 1.57 \\
2.40 \quad & 3.91 \quad & 4.58 \quad &  7.50 \quad & 0.08 \quad & 0.30 \quad & 0.38 \quad & 1.80 \\
2.50 \quad & 4.50 \quad & 3.80 \quad &  7.50 \quad & 0.08 \quad & 0.30 \quad & 0.37 \quad & 1.80 \\
2.70 \quad & 5.00 \quad & 3.45 \quad &  7.80 \quad & 0.08 \quad & 0.31 \quad & 0.39 \quad & 1.87 \\
3.00 \quad & 4.50 \quad & 3.50 \quad &  8.10 \quad & 0.08 \quad & 0.31 \quad & 0.38 \quad & 1.94 \\
3.25 \quad & 4.50 \quad & 3.50 \quad &  8.10 \quad & 0.08 \quad & 0.32 \quad & 0.38 \quad & 1.93 \\
\end{tabular}
\caption{Optimized material parameters ({MP}) such as shell elastic constants $\left(C_1,C_2\right)\,$MPa, bubble shell thickness $(h)\,$nm, viscosity of shell $(\eta^{\rm S})\,$Pa-s,  natural configuration pressure $(p_{g_0})\,$MPa, shell elasticity modulus $(\chi)\,$N/m and shell dilatational viscosity $(k^{\rm {S}}\,)10^{-9}\,$kg/s for encapsulated bubbles with different natural configuration outer radii $(R_{20})\,\mu$m, viscosity of the liquid $\eta^{\rm L}=1\,$mPa-s, excitation pressure $p_a=0.15\,$MPa and frequency $f=2.5\,$MHz.}
\label{stb:2}
\end{center}
\end{table}